\documentclass[aps,prl,twocolumn,superscriptaddress,allowtoday]{revtex4-2}
\pdfoutput=1

\usepackage[utf8]{inputenc}
\usepackage[english]{babel}
\usepackage[a4paper, margin=1in]{geometry}
\usepackage{graphicx} 
\usepackage{amsmath}
\usepackage{amsthm}
\usepackage{amsfonts}
\usepackage{physics}
\usepackage{hyperref}
\usepackage{cleveref}
\usepackage{placeins}
\usepackage{tikz}
\usepackage[normalem]{ulem}

\newcommand{\prlsection}[1]{\emph{#1}---}

\begin{document}

\title{Bacon-Shor Board Games}

\author{M. Sohaib Alam}
\email{malam@usra.edu}
\affiliation{These authors contributed equally}
\affiliation{Quantum Artificial Intelligence Laboratory (QuAIL), NASA Ames Research Center, Moffett Field, CA, 94035, USA}
\affiliation{USRA Research Institute for Advanced Computer Science (RIACS), Mountain View, CA, 94043, USA}

\author{Jun Zen}
\email{jun.j.zen@gmail.com}
\affiliation{These authors contributed equally}
\affiliation{Okinawa Institute of Science and Technology, Okinawa, 904-0495, Japan}

\author{Thomas R. Scruby}
\email{t.r.scruby@gmail.com}
\affiliation{Okinawa Institute of Science and Technology, Okinawa, 904-0495, Japan}

\date{\today}

\begin{abstract}
    We identify a period-4 measurement schedule for the checks of the Bacon-Shor code that fully covers spacetime with constant-weight detectors, and is numerically observed to provide the code with a threshold. Unlike previous approaches, our method does not rely on code concatenation and instead arises as the solution to a coloring game on a square grid. Under a uniform circuit-level noise model, we observe a threshold of approximately $0.3\%$ when decoding with minimum weight perfect matching.
\end{abstract}

\maketitle

\prlsection{Introduction}
Quantum computing holds the potential of revolutionizing certain computational tasks that are intractable for classical hardware, ranging from simulation of physical systems \cite{lloyd1996universal,berry2007efficient,childs2012hamiltonian,childs2017quantum,low2019hamiltonian,babbush2018encoding,bauer2023quantum,shaw2020quantum,jordan2012quantum,hardy2024optimized},
cryptography \cite{shor1994algorithms,wiesner1983conjugate,bennett2014quantum,ekert1991quantum,bennett1992quantum,shor2000simple,mayers2001unconditional,lo1999unconditional,gisin2002quantum,acin2006bell,pirandola2020advances}
and optimization problems \cite{jordan2024optimization,farhi2001quantum,farhi2014quantum,hadfield2019quantum,alam2022practical,dalzell2023mind,kapit2023approximability,shaydulin2024evidence,pirnay2024principle,szegedy2022quantum}. 
However it is hindered by errors arising from decoherence and imperfect control \cite{krantz2019quantum}. Quantum error correction (QEC) is therefore essential for scalable quantum computation.

The Bacon-Shor code \cite{bacon2006operator} is a prototypical quantum error correcting subsystem code defined on a square lattice with qubits residing on the vertices, requiring only nearest-neighbor, weight-2 geometrically local measurements.
These properties
make the code attractive from an experimental perspective, enabling fault-tolerance \cite{aliferis2007subsystem,yoder2017universal} and early demonstrations on nascent quantum hardware \cite{egan2021fault,li2018direct,huang2024comparing}. However, the stabilizers of the Bacon-Shor code grow in weight as $O(d)$ on a $d \times d$ lattice.
As $d$ increases, the growing weight of stabilizer measurements reduces their reliability, causing the logical error rate to eventually exceed the physical error rate, thereby preventing the code from having a threshold.


Recent work \cite{gidney2023baconthreshold} addressed this by concatenating Bacon–Shor patches via lattice surgery to reduce logical error rates. Their construction effectively reduces detector weight by using only a subset of the gates from a larger non-concatenated Bacon–Shor circuit. This “gate deletion” strategy avoids large size detectors while maintaining fault-tolerance, and enables logical error suppression beyond the capabilities of the standard Bacon–Shor code. However, the resultant measurement schedule results in both the maximum weight of any detector as well as the period of the full measurement cycle scaling as $\Omega(d)$ on a $d \times d$ square lattice.

Since the growth of the weight of the detectors as $O(d)$ in the Bacon-Shor code with the usual measurement schedule is ultimately the reason why the code lacks a threshold, motivated by the observation of \cite{gidney2023baconthreshold} that an alternative measurement schedule can fix this issue, we ask whether a measurement schedule is possible such that all detectors have weight that grow less than $O(d)$? In this paper, we show that not only is this possible, but that this weight can be reduced to $O(1)$, and that it is possible to achieve this with a measurement cycle of $O(1)$ period. 
Additionally, in numerical simulations of this measurement schedule under a circuit-level noise model we observe a threshold, as well as logical error rates significantly lower than all previously known measurement schedules for this code.


As with other subsystem codes, the Bacon-Shor code can be specified via its gauge group, which in turn is in general a non-abelian subgroup of the Pauli group. In this case, the generators of the gauge group are defined on the edges of a square lattice with qubits residing on the nodes of this lattice. Horizontal edges give rise to nearest neighbor $XX$ operators, while vertical edges correspond similarly to $ZZ$ operators. These `check operators' generate the full gauge group, and correspond to the set of operators one would measure. 

On a $d \times d$ square lattice, we obtain a gauge group with $2d(d-1)$ independent generators, corresponding to the number of edges on the square lattice. The center of this gauge group is a subgroup consisting of all elements in the gauge group that commute with every element of the gauge group, and is identified with the stabilizer subgroup. There are $2(d-1)$ independent stabilizer generators, each corresponding to the product of $X$ ($Z$) type checks along neighboring columns (rows). Each of these stabilizers is of weight $2d$. Since there are a total of $d^2$ physical qubits, we are left with $d^2 - 2(d-1) = (d-1)^2 + 1$ logical qubits. Only one of these has logical operators that cannot be written as products of the measured two-qubit check operators, so only information stored in this single logical qubit is protected while all other so-called ``gauge qubits'' are measured out, or ``gauge fixed'', by the physical measurements. Each of these gauge qubits can be associated to a square plaquette of the lattice, and its logical $\overline{X}$ or $\overline{Z}$ operator can be written as the product of all $X$ checks above or all $Z$ checks to the right of this plaquette respectively~\cite{poulin2005stabilizer,li20192d,alam2025dynamical}.

A typical measurement schedule for the Bacon-Shor code, and indeed for any CSS subsystem code, is a period-2 cycle in which we measure all $X$ type checks, then measure all $Z$ type checks, and then repeat. Such a measurement prescription only allows for the $O(d)$ weight stabilizers to form detectors, and causes this code to lack a threshold as we grow $d$, the size of the lattice. Below, we describe how we form detectors out of other combinations of gauge check operators, by following an alternative measurement schedule.

\begin{figure*}
    \centering
\includegraphics[width=\linewidth]{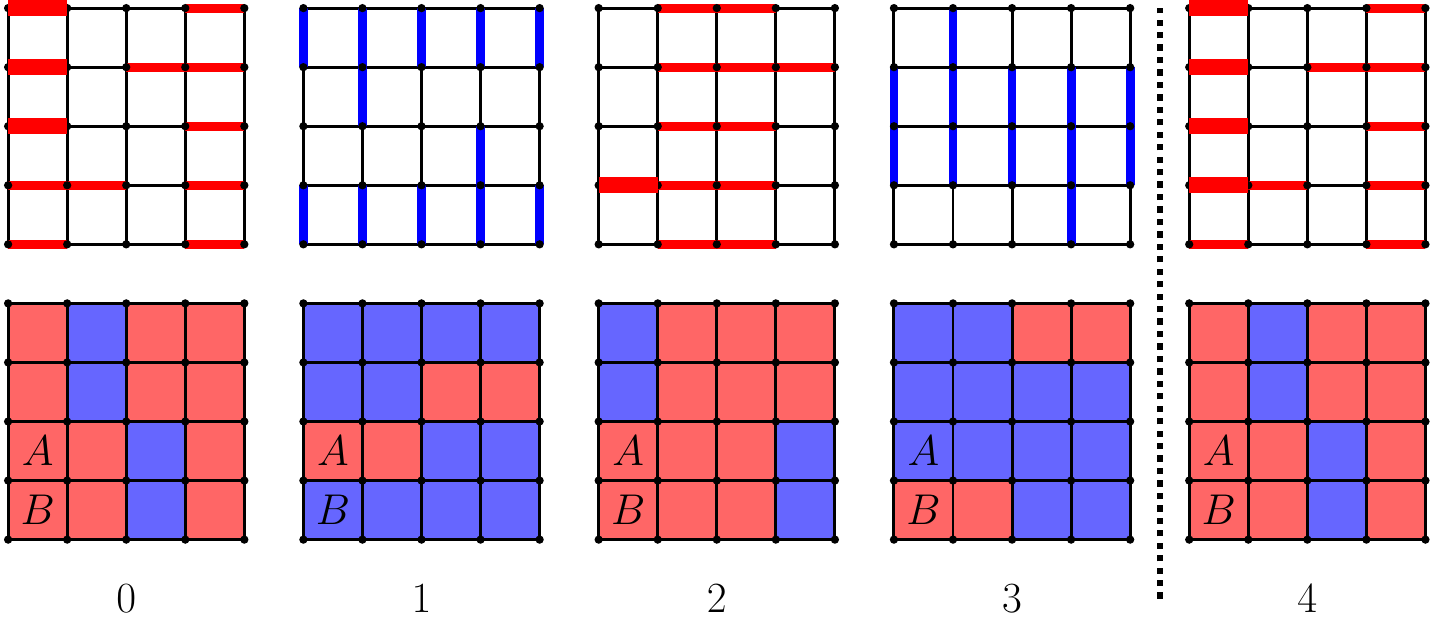}
\caption{(Top) Period $4$ measurement schedule on a $5 \times 5$ lattice, where horizontal $XX$ checks are drawn in red, and vertical $ZZ$ checks in blue. (Bottom) Induced ISG sequence after one complete measurement cycle. This periodic sequence of colorings solves the game described in the main text. Measured checks contributing to an example detector are drawn in thicker red lines compared to the rest.}
\label{fig:meas-ISGs}
\end{figure*}

\prlsection{Measurement schedule}
We obtain a period-4 measurement schedule, shown in Fig.~\ref{fig:meas-ISGs}, as a solution to a coloring game on a square lattice.
The coloring game we use to obtain our measurement schedules can be understood as a sequence of gauge fixings for the gauge qubits of the Bacon-Shor code.

We can now introduce this game, which has the following objective: given a $d \times d$ square grid, find a periodic sequence of colorings of the plaquettes of this grid where each plaquette is colored either red or blue and
\begin{enumerate}
    \item each column (row) is colored entirely red (blue) in at least one coloring;
    \item each column (row) contains at least one red (blue) plaquette in each coloring;
    \item the period of the sequence is as short as possible.
\end{enumerate}
The rules above ensure that
\begin{enumerate}
    \item there is at least one measurement round where we measure all the $X$ ($Z$) checks in a given column (row), meaning we can initialize and measure all detectors;
    \item products of $X$ ($Z$) checks with lesser weight than the Bacon-Shor code stabilizers in the same column (row) are preserved in each round, meaning detectors that are spatially small can be built from these products;
    \item the period of the full measurement cycle is short, and so the resulting detectors are small in time.
\end{enumerate}
Additionally, we do not allow arbitrary sequences and instead each coloring in the sequence must be obtained from the previous one using the following move: if in the current coloring, a box is painted red (blue), then in the subsequent coloring, a vertical (horizontal) strip of any length may be colored red (blue) as long as it contains the original red (blue) box from the previous coloring. 
For the purposes of the coloring game, we allow strips with periodic boundary conditions, so that e.g. if the bottom box in a column is colored red, then the top box in the same column may also be colored red in the subsequent coloring as long as the bottom box continues to be colored red in the subsequent coloring as well.
Note that while we allow periodic boundary conditions for colored strips, the underlying Bacon-Shor code continues to be embedded on a 2D surface with open (non-periodic) boundary conditions.

The coloring sequence of Fig. \ref{fig:meas-ISGs} solves such a board game for a $5 \times 5$ lattice.
An example of a detector formed along the left most plaquette column from this measurement schedule is given by the product $\overline{X}_A^{(0)} \cdot \overline{X}_A^{(2)} \overline{X}_B^{(2)} \cdot \overline{X}_B^{(4)}$, where $\overline{X}_A^{(r)}$ $\left(\overline{X}_B^{(r)}\right)$ denotes the measured product of all the $XX$ checks above cell $A$ ($B$), or alternatively the logical $\overline{X}$ operator of gauge qubit A (B), at round $r$. The detector can also be thought of as the product of all the measured checks highlighted with thicker red lines than the other measured checks shown in Fig. \ref{fig:meas-ISGs}. Note that the product $\overline{X}_A^{(2)} \overline{X}_B^{(2)}$ equals the $XX$ check straddling cells $A$ and $B$ at round $2$, and round $4$ denotes round $0$ of the subsequent measurement cycle.


We would ideally like to solve the board game for as small a value of $d$ and period $R$ as possible. The reason is that
once a solution is found for any finitely sized board of length $d$ for a finite period $R$, resulting in finitely sized detectors of weight $O(Rd)$, one can patch these together to form solutions for arbitrary sized lattices with the same period, and similarly finitely sized detectors of weight $O(Rd)$, as depicted in Fig. \ref{fig:meas-ISGs-larger}, and explained in more detail in the Supplementary Material.
An example detector cutting across patches obtained from Fig. \ref{fig:meas-ISGs} is depicted in Fig. \ref{fig:meas-ISGs-larger}, and is given by the product of highlighted red $X$ checks, or in terms of the gauge qubit logical operators, $\left( \overline{X}_{A}^{(0)} \overline{X}_{C}^{(0)} \right) \left( \overline{X}_{A}^{(2)} \overline{X}_{B}^{(2)} \cdot \overline{X}_{C}^{(2)} \overline{X}_{D}^{(2)}\right) \left( \overline{X}_{B}^{(4)} \overline{X}_{D}^{(4)} \right)$, where $\overline{X}_g^{(t)}$ denotes the logical $X$ operator of gauge qubit $g$ measured at round $t$. Such a detector always consists of the product of 10 checks, each of weight 2, so its total weight (20) remains a constant regardless of how many smaller sized patches are put together.

We note that while solutions to the board game on a $3 \times 3$ lattice can be ruled out, a solution to a $4 \times 4$ lattice with period 6 can be constructed. For simplicity, we choose to focus on the $5 \times 5$ solution with a smaller period of 4. Details on smaller sized lattices are provided in the Supplementary material.

\begin{figure*}[t!]
    \centering
\includegraphics[width=\linewidth]{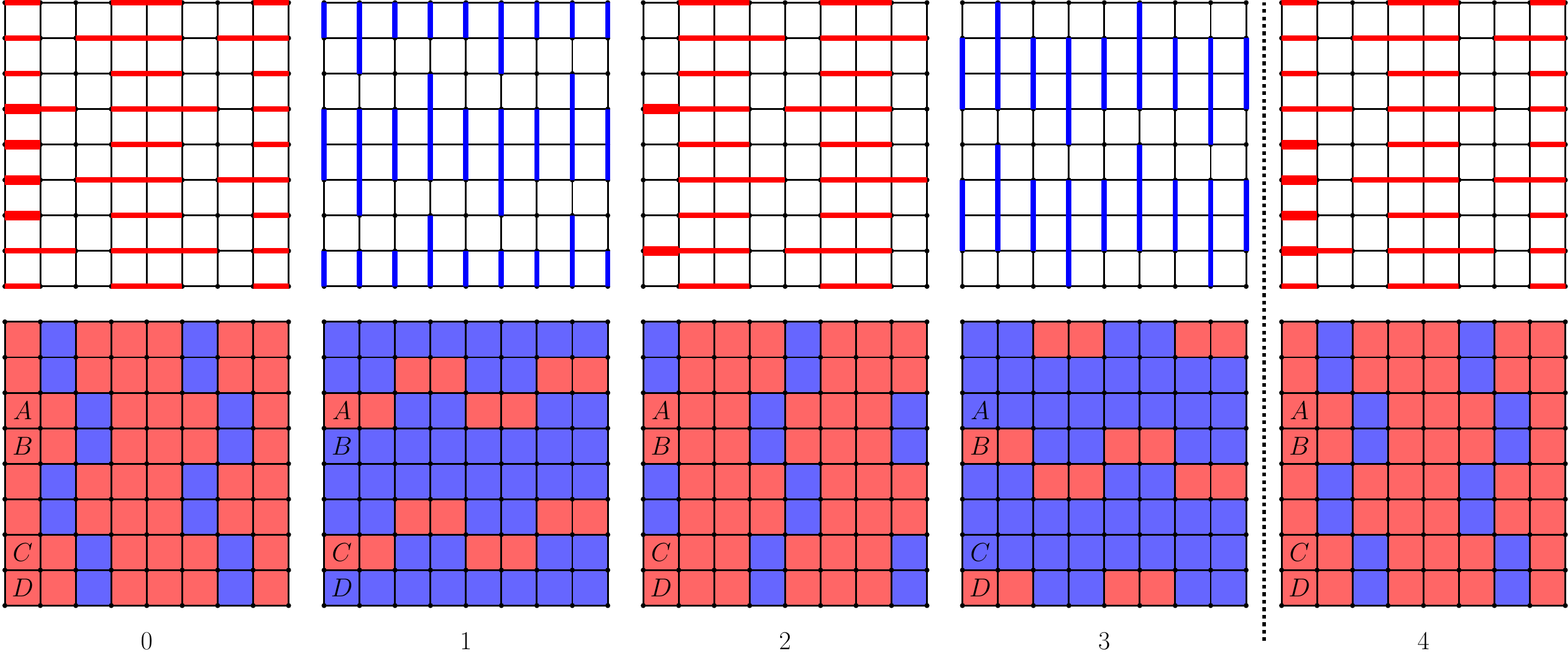}
\caption{Four copies of the smaller sized solution of Fig. \ref{fig:meas-ISGs} patched together to form a larger lattice sized solution to the board game described in the main text. Measured checks contributing to an example detector that cuts across patches obtained from the solution in Fig. \ref{fig:meas-ISGs} are drawn in thicker red lines compared to the other $X$ type measurements.}
\label{fig:meas-ISGs-larger}
\end{figure*}

\prlsection{Numerical results}
We evaluate the performance of the modified Bacon–Shor code measurement schedule in Fig. \ref{fig:meas-ISGs} under a circuit-level noise model using \texttt{Stim}~\cite{gidney2021stim}. For consistency with the convention of~\cite{gidney2023baconthreshold}, we define one ``round'' as two steps out of our four-step cycle, allowing for a direct comparison of logical error rates with that work since they also define a ``round'' in the same way. All circuits (initialized in the $X$ basis for these tests) are simulated using a uniform circuit-level noise model defined in the 
Supplementary Material,
where every physical operation (single- or two-qubit gate, measurement, or reset) fails independently with probability $p$. All decoding is performed using PyMatching~\cite{higgott2021pymatching,higgott2025sparse}. 

Following the approach in \cite{gidney2023baconthreshold}, we stop each data collection once we have sampled $10^8$ shots or observed $10^3$ logical errors, whichever comes first. This ensures each reported logical error probability has sufficient statistical accuracy.

\begin{figure}[t!]
    \centering
    \includegraphics[width=1\linewidth]{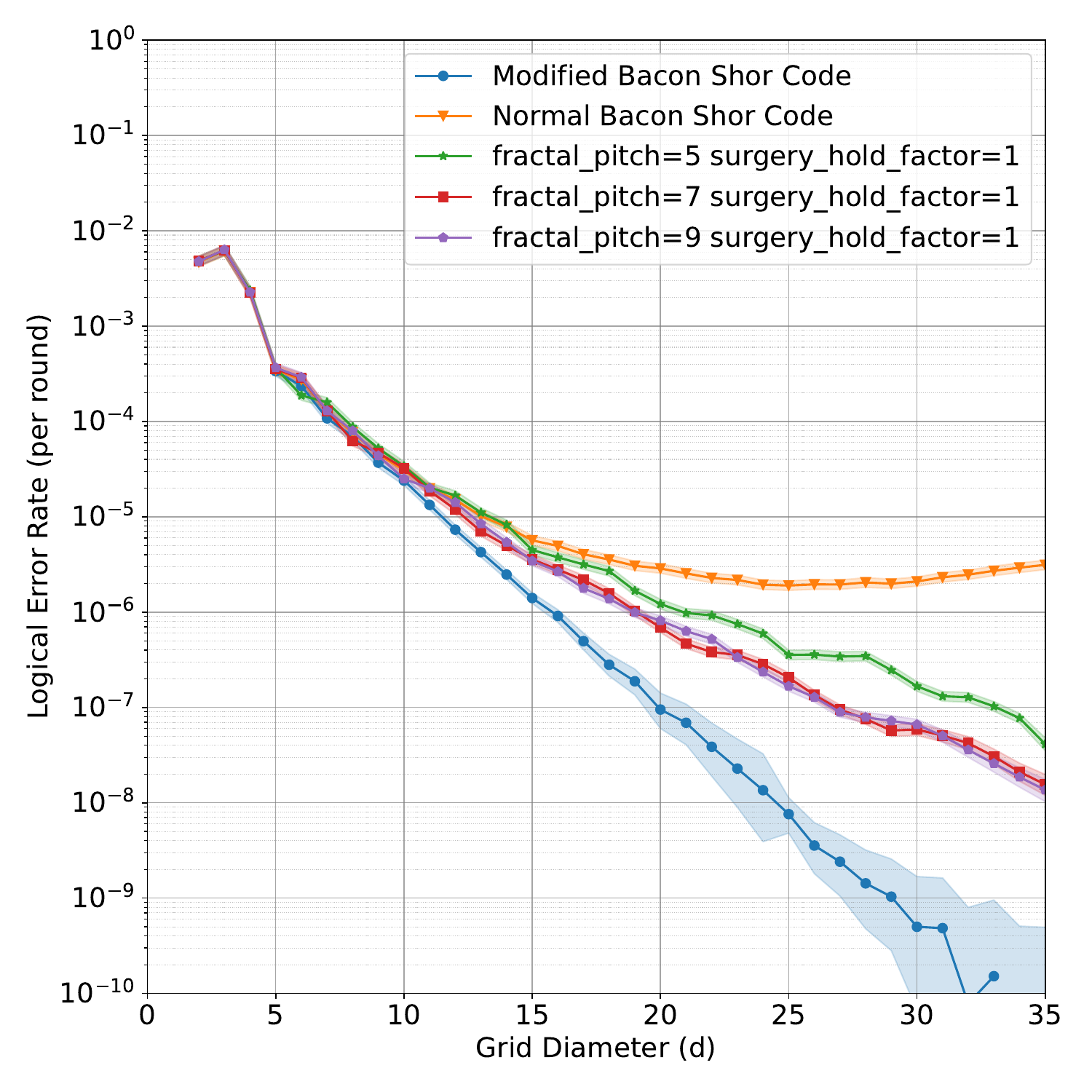}
    \caption{Comparison of the per-round logical $Z$ error rate for three implementations of the Bacon-Shor code, our modified version (blue), the standard (unmodified) code (orange), and the fractal-pitch variants from \cite{gidney2023baconthreshold} (pitches 5, 7, 9 with surgery hold factor = 1). We simulate $4d$ measurement rounds with a noise strength of $p=0.001$ for in each case. Shaded regions represent statistical confidence intervals.}
    \label{fig:error-rate}
\end{figure}

In Fig.~\ref{fig:error-rate} we plot logical error rate per round (for one round as defined above) vs lattice size for various measurement schedules and observe that our modified schedule achieves an exponentially decreasing logical error rate with increasing $d$. In contrast, the unmodified Bacon–Shor code eventually saturates or worsens at larger distances. Notably, our logical error rates are two orders of magnitude lower than those from the self-concatenation approach of \cite{gidney2023baconthreshold}. This clear exponential suppression of logical error rate with code size indicates that $p=10^{-3}$ is below the threshold of the code. 

\begin{figure}[t!]
    \centering
    \includegraphics[width=1\linewidth]{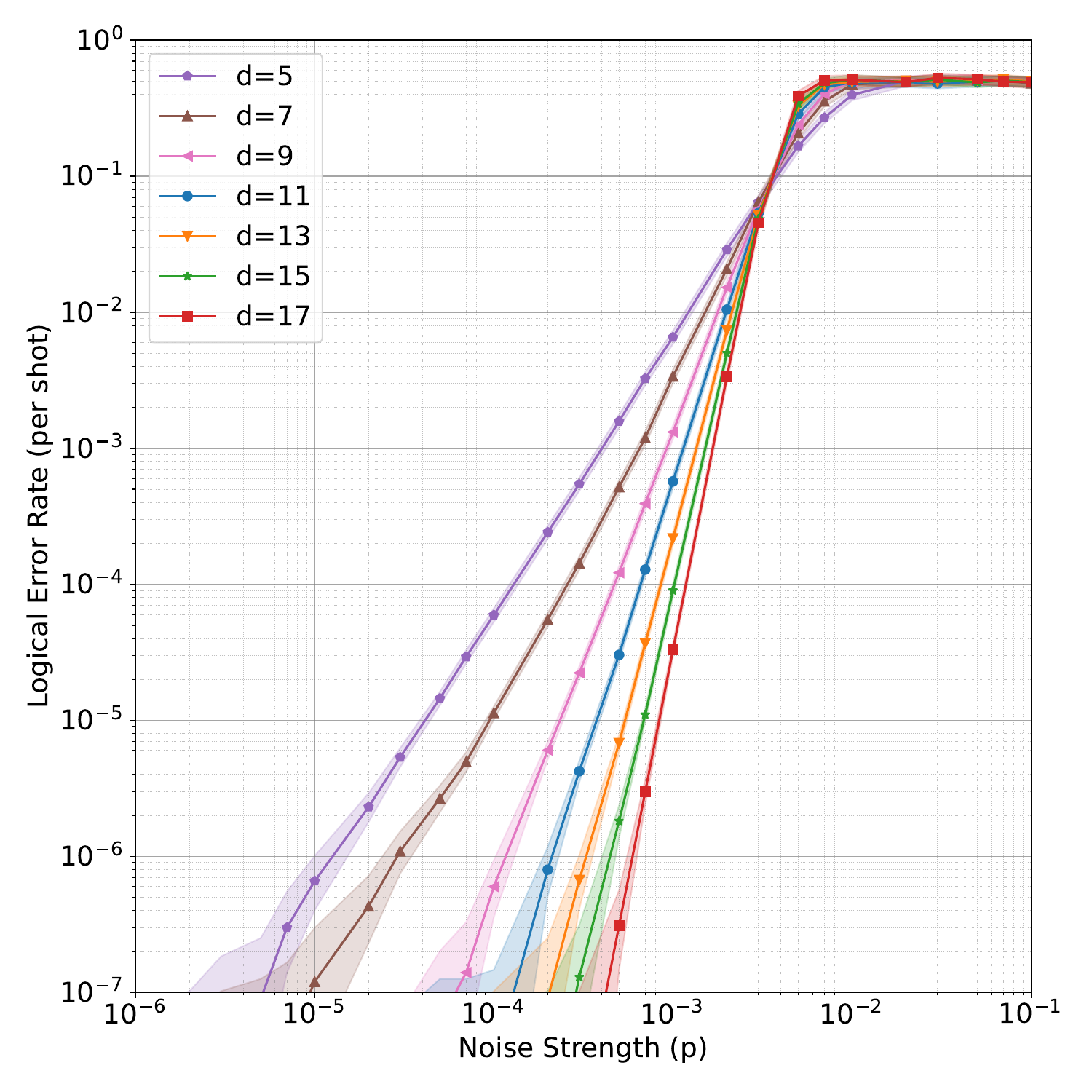}
    \caption{Physical versus logical (per-shot) error rate in the Bacon-Shor code using the modified check measurement schedule from Fig. \ref{fig:meas-ISGs}. $4d$ measurement rounds are simulated for each shot.}
    \label{fig:threshold}
\end{figure}

To find an approximate value of this threshold we simulate the performance of the code for different grid diameters $d$ with increasing noise strength as shown in Fig.~\ref{fig:threshold}, and observe a threshold of around $p_{\text{th}} \approx 3 \times 10^{-3}$.   

\label{secn:num-results}

\prlsection{Conclusion}
We have shown that the Bacon–Shor code acquires a threshold via a minor modification of its standard measurement schedule. By exploiting the subsystem structure through a specific choice of virtual qubit operators, our approach enables constant weight detectors and achieves a threshold competitive with the surface code \cite{fowler2012surface}, one that is moreover within the noise levels of current hardware \cite{acharya2024quantum,paetznick2024demonstration}. Similar ideas were used in \cite{alam2025dynamical} to increase the encoding rate by introducing dynamical logical qubits, saturating the subsystem BPT bound \cite{bravyi2011subsystem,bravyi2009no,bravyi2010tradeoffs,baspin2024improved,dai2024locality}.

Our work highlights the importance of the measurement schedule in determining crucial properties of subsystem codes. In particular, the dynamical aspect of various measurement schedules can produce completely different properties using the same set of measured check operators. This result emphasizes the need to identify a broader set of unifying principles for quantum error correction, one that includes the order in which checks of the code are measured at a fundamental level. We find it noteworthy that the approach used in this work directly exploits the subsystem structure of the Bacon-Shor code to reveal the shortcomings of the usual formalism to describe subsystem codes. We anticipate future work would shed greater light on the algebraic structure and temporal aspects of quantum error correcting codes, and reveal a better suited formalism to unify several such ideas currently being explored in the literature.

\prlsection{Acknowledgements}
We would like to thank David Elkouss, Oscar Higgott, Joan Camps, Eleanor Rieffel, and Jason Saied for helpful conversations. Part of this work was initiated when M.S.A. and T.R.S. were visiting the Fault-Tolerant Quantum Technologies Workshop in Benasque, 2024. We are grateful to the organisers of the workshop.

M.S.A. acknowledges support from USRA NASA Academic Mission Services under contract No. NNA16BD14C with NASA, and from the U.S. Department of Energy, Office of Science, National Quantum Information Science Research Centers, Superconducting Quantum Materials and Systems (SQMS) Center under contract No. DEAC02-07CH11359 and Co-Design Center for Quantum Advantage (C2QA) under Contract No. DESC0012704.
J.Z. acknowledges support from the JST Moonshot R$\&$D
program under Grant JPMJMS226C.
T.R.S. was supported by the JST Moonshot R\&D Grant [grant number JPMJMS2061] for part of this work. 
Numerical results presented in this work were obtained using the HPC resources provided by the Scientific Computing and Data Analysis section of the Research Support Division at OIST. 

AI tools were used to assist with writing parts of the code and improving grammatical correctness and formatting of the text. All AI generated content was reviewed by the authors, who take full responsibility for the final form of the content.  


\bibliography{references}

\end{document}


\title{Supplemental material for the paper ``Bacon-Shor Board Games"}

\author{M. Sohaib Alam}
\email{malam@usra.edu}
\affiliation{These authors contributed equally}
\affiliation{Quantum Artificial Intelligence Laboratory (QuAIL), NASA Ames Research Center, Moffett Field, CA, 94035, USA}
\affiliation{USRA Research Institute for Advanced Computer Science (RIACS), Mountain View, CA, 94043, USA}

\author{Jun Zen}
\email{jiajun.chen@oist.jp}
\affiliation{These authors contributed equally}
\affiliation{Okinawa Institute of Science and Technology, Okinawa, 904-0495, Japan}

\author{Thomas R. Scruby}
\email{t.r.scruby@gmail.com}
\affiliation{Okinawa Institute of Science and Technology, Okinawa, 904-0495, Japan}

\date{\today}

\maketitle

\section{Canonical Basis}
In general, any subsystem code is specified by its gauge group, which is in general a non-abelian subgroup of the Pauli group, defined over $n$ qubits as $\mathcal{P}_n = \langle i\mathbb{I}, X_1, Z_1, \dots, X_n, Z_n \rangle$. The gauge group $\mathcal{G} \subset \mathcal{P}_n$ is typically generated by low weight ``check" operators that do not necessarily commute with one another. The center of this gauge group is defined as $\mathcal{Z}(\mathcal{G}) = C(\mathcal{G}) \cap \mathcal{G}$, where $C(\mathcal{G}) = \lbrace h \in \mathcal{P}_n \vert [h,g]=0 \;\;\forall g \in \mathcal{G} \rbrace$ is the centralizer. The center consists of all elements of the gauge group that commute with every element of the gauge group, and is identified as the stabilizer subgroup $\mathcal{Z}(\mathcal{G}) = \mathcal{S}$. Its elements are typically the ones that form detector syndromes used for error correction. Typically, the weights of the stabilizer generators are larger than those of the gauge group generators, allowing one to extract the value of higher weight operators from the measurement of smaller weight check operators. Bare logical operators are defined to be elements of $\mathcal{L}_b = C(\mathcal{G}) \backslash \mathcal{G}$, while dressed logical operators are defined as $\mathcal{L}_d = C(\mathcal{S}) \backslash \mathcal{G} \simeq \mathcal{G} \mathcal{L}_b$.

Using a symplectic Gram-Schmidt method \cite{wilde2009logical}, the gauge group can generically be written in canonical form as $\mathcal{G} = \langle \overline{Z}_1, \dots, \overline{Z}_{s}, \overline{X}_{s + 1}, \overline{Z}_{s + 1}, \dots, \overline{X}_{s + g}, \overline{Z}_{s + g}\rangle$, where $s$ denotes the rank of the stabilizer subgroup, and $\vert G \vert = s + 2g$ provides the rank of the gauge group. We may think of the sets $\{ \overline{Z}_{i}\}_{i=1}^{s}$ as the logical $\overline{Z}$ and $\{\overline{X}_j, \overline{Z}_j \}_{j=s + 1}^{s + g}$ as the logical operators of ``virtual" stabilizer and gauge qubits, respectively. These virtual qubits are collective degrees of freedom over the physical qubits, and provide a convenient manner with which to reason about error correcting codes. Indeed, the entire Pauli group can be written, under an automorphism, as
\begin{eqnarray}
\mathcal{P}_n &=& \langle i\mathbb{I}, \overline{X}_1, \overline{Z}_1, \dots, \overline{X}_{s}, \overline{Z}_{s}, \nonumber \\
&& \overline{X}_{s + 1}, \overline{Z}_{s + 1}, \dots, \overline{X}_{s + g}, \overline{Z}_{s + g}, \nonumber \\
&& \overline{X}_{s + g + 1}, \overline{Z}_{s + g + 1}, \dots, \overline{X}_{s + g + k}, \overline{Z}_{s + g + k} \rangle
\end{eqnarray}
for a subsytem code with $s$ stabilizer qubits, $g$ gauge qubits, and $k$ logical qubits, such that $s + g + k = n$, where $n$ is the total number of physical qubits. Note that instead of fixing the value of all $\overline{Z}$s for the stabilizer qubits, we could also have instead identified the corresponding fixed stabilizer generator as the $\overline{X}$ operator of some stabilizer qubit. We adopt a convention of fixing respectively the $\overline{Z}$ and $\overline{X}$ operators for the `horizontal' and `vertical' stabilizer qubits in our particular case. 

On a $d \times d$ lattice, the gauge group of the Bacon-Shor code admits $d-1$ $X$-type (vertical) stabilizers, and $d-1$ $Z$-type (horizontal) stabilizers, thus giving rise to $2(d-1)$ `stabilizer qubits'. There are also $(d-1)^2$ `gauge qubits', leaving room for $1$ logical qubit. Each of the stabilizer, gauge and logical qubits are virtual qubits, i.e. collective qubit-like degrees of freedom over several physical qubits.

Formally, we can write the gauge group for the Bacon-Shor code defined on a $d \times d$ lattice as $\mathcal{G} = \left\langle X_{i,j} X_{i,j+1}, Z_{i,j}Z_{i+1,j} \mid i,j \in [d] \right\rangle$, where $[d] = \lbrace 0, \dots, d-1 \rbrace$, and the origin lies at the bottom-left corner of the lattice. In the canonical basis, this gauge group can be rewritten in terms of stabilizer and gauge qubit operators as
\begin{eqnarray}
\mathcal{G} &=& \left\langle \overline{Z}_1, \dots, \overline{Z}_{d-1}, \overline{X}_{d}, \dots, \overline{X}_{2(d-1)}\right. \nonumber \\
&& \overline{X}_{2(d-1) + 1}, \overline{Z}_{2(d-1) + 1}, \dots, \nonumber \\
&& \left. \overline{X}_{2(d-1) + (d-1)^2}, \overline{Z}_{2(d-1) + (d-1)^2} \right\rangle
\label{eqn:gauge-group-canonical}
\end{eqnarray}
where the logical $\overline{Z}$ and $\overline{X}$ operators of the horizontal and vertical stabilizer qubits respectively are given by
\begin{widetext}
\begin{equation}
\begin{alignedat}{5}
&\text{Horizontal $Z$-type stabilizers:} &\qquad& \overline{Z}_i = \prod_{j=0}^{d-1} Z_{j,i} Z_{j,i-1} &\qquad & i \in \lbrace 1,\dots,d-1 \rbrace &\qquad & \\
&\text{Vertical $X$-type stabilizers:} &\qquad& \overline{X}_{d+j} = \prod_{k=0}^{d-1} X_{j,k} X_{j+1,k} &\qquad & j \in \lbrace 0,\dots,d-2 \rbrace &\qquad &
\end{alignedat}
\label{eqn:canonical-stabilizer-ops}
\end{equation}
\end{widetext}
There are now $(d-1)^2$ remaining gauge qubit degrees of freedom, precisely equaling the number of square plaquettes on a $d \times d$ lattice. In Eq.~\eqref{eqn:gauge-group-canonical}, the virtual qubit operators with indices $2(d-1)+1,\dots,2(d-1)+(d-1)^2$ act on each of these $(d-1)^2$ many gauge qubits. We associate gauge qubit $q \in {1,\dots,(d-1)^2}$ with the square plaquette whose bottom-left corner is at coordinate $(x_q,y_q)$ where
\begin{eqnarray}
x_q &=& \left\lfloor \frac{q-1}{d-1} \right\rfloor \nonumber \\
y_q &=& (q-1)\; \text{mod}\; (d-1)
\end{eqnarray}
with the origin assumed to be at the bottom-left corner of the lattice. With this identification, we can write the logical operators of the gauge qubits as
\begin{eqnarray}
\overline{X}_{i,j} &=& \prod_{k=j+1}^{d-1} X_{i,k} X_{i+1,k} \nonumber \\
\overline{Z}_{i,j} &=& \prod_{l=i+1}^{d-1} Z_{l,j} Z_{l,j+1}
\label{eqn:gauge-qubit-logicals}
\end{eqnarray}
where $i,j \in \lbrace 0,\dots,d-1 \rbrace$. The operators given in Eqs.~\eqref{eqn:canonical-stabilizer-ops} and \eqref{eqn:gauge-qubit-logicals} generate the entire gauge group. A choice of bare logical operators is then
\begin{eqnarray}
\overline{X}_{\text{logical}} &=& \prod_{j=0}^{d-1} X_{0,j} \nonumber \\
\overline{Z}_{\text{logical}} &=& \prod_{j=0}^{d-1} Z_{j,0}
\label{eqn:bacon-shor-logicals}
\end{eqnarray}
Finally, we can also make a choice of ``destabilizers", Pauli operators that act as the logical $\overline{X}$ and $\overline{Z}$ operators of horizontal and vertical stabilizer qubits respectively.
\begin{widetext}
\begin{equation}
\begin{alignedat}{5}
&\text{Destabilizers for horizontal $Z$-type stabilizers:} &\qquad& \overline{X}_i = \prod_{j=i}^{d-1} X_{0,j}&\qquad & i \in \lbrace 1,\dots,d-1 \rbrace &\qquad & \\
&\text{Destabilizers for vertical $X$-type stabilizers:} &\qquad& \overline{Z}_{d+j} = \prod_{k=j+1}^{d-1} Z_{k,0} &\qquad & j \in \lbrace 0,\dots,d-2 \rbrace &\qquad &
\end{alignedat}
\label{eqn:canonical-destabilizer-ops}
\end{equation}
\end{widetext}
Note that the destabilizers for horizontal (respectively, vertical) stabilizers are given by a chain of vertical (respectively, horizontal) Pauli operators.

\begin{figure*}
    \centering
\includegraphics[width=\linewidth]{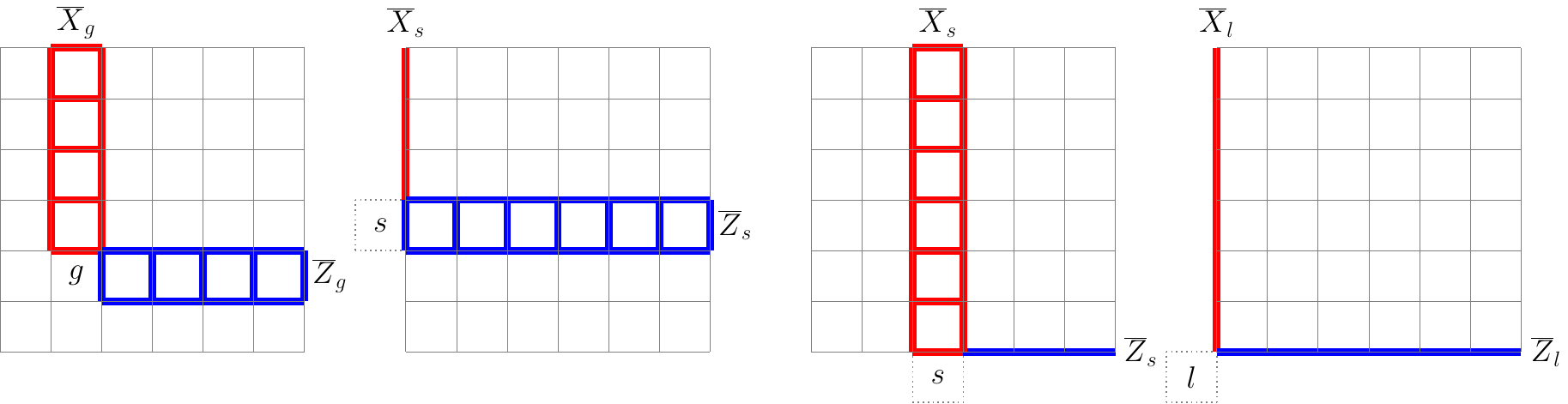}
\caption{Virtual qubit operators that generate the entire Pauli group. L-R: an example of a pair of logical operators of each of the $(d-1)^2$ gauge qubits in the Bacon-Shor code; an example of the logical $\overline{Z}$ operator of a `horizontal' stabilizer qubit, along with its destabilizer; an example of the logical $\overline{X}$ operator of a `vertical' stabilizer qubit, along with its destabilizer; a representative pair of the $\overline{X}$ and $\overline{Z}$ operators of the single logical qubit in the Bacon-Shor code.}
\label{fig:virtual-qubit-operators}
\end{figure*}

The operators given by Eqs.~\eqref{eqn:canonical-stabilizer-ops}, \eqref{eqn:gauge-qubit-logicals},, \eqref{eqn:bacon-shor-logicals}, and \eqref{eqn:canonical-destabilizer-ops}, along with the operator $i\mathbb{I}$, generate the Pauli group over $d^2$ qubits. All generators of the Pauli group in this basis are depicted in Fig. \ref{fig:virtual-qubit-operators}.

\section{Allowed moves}
\label{appdx:game}

\begin{figure*}[!t]
    \centering
\includegraphics[width=\linewidth]{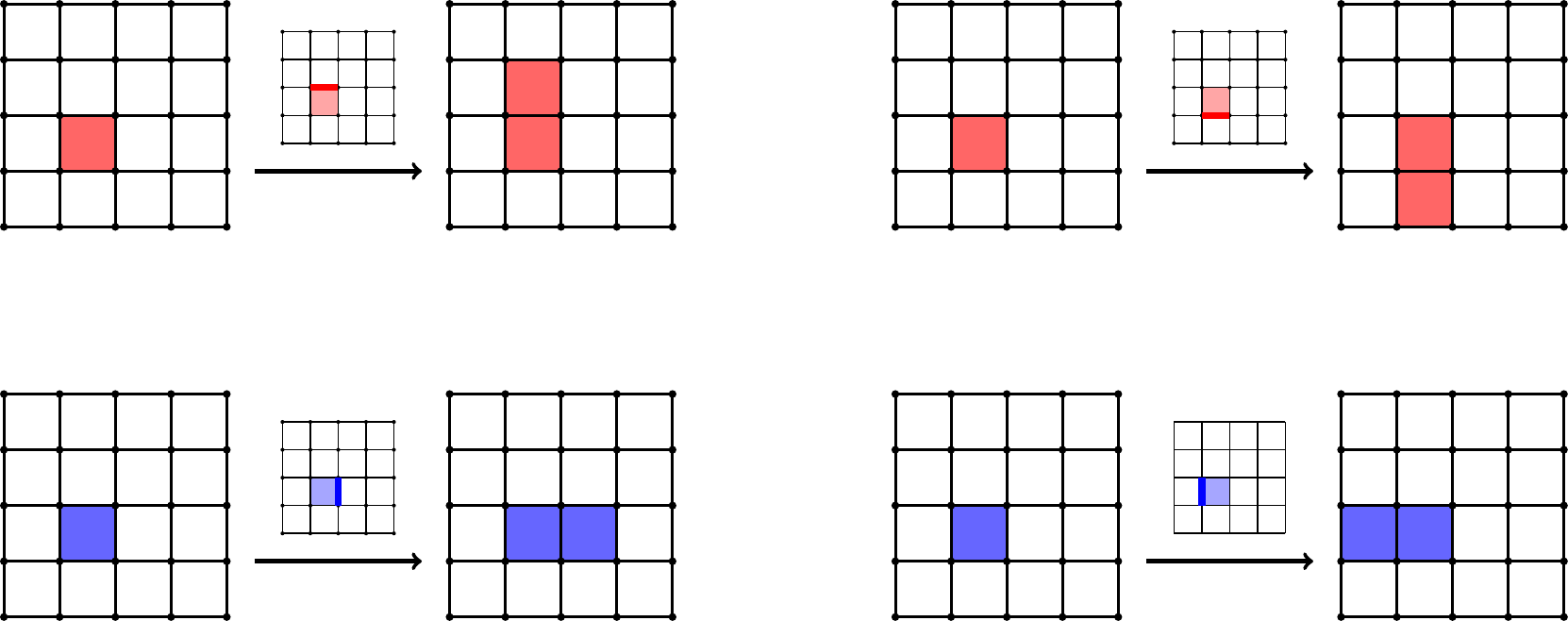}
\caption{Basic moves for propagating gauge fixed operators. Measuring $XX$ or $ZZ$ checks propagates red or blue boxes respectively. Vertical or horizontal strips of arbitrary length can be colored red or blue in this manner by measuring several $XX$ or $ZZ$ checks. This gives rise to moves (i) and (ii) described in the main text.}
\label{fig:basic-moves}
\end{figure*}

The moves used to progress from one coloring to the next in a given sequence directly translate from measuring the checks of the Bacon-Shor code. The basic check measurements that give rise to the fundamental type of allowed move described in the main text are illustrated in Fig. \ref{fig:basic-moves}. As argued in the main text, a box, or square plaquette, is associated with a gauge qubit of the Bacon-Shor code. Coloring this box red (\red) or blue (\blue) corresponds to fixing the value of the associated gauge qubit's $\overline{X}$ or $\overline{Z}$ operator, respectively. If there exists a gauge fixed $\overline{X}$ operator for some box, measuring an $XX$ check above this box fixes the $\overline{X}$ operator of the box directly above it as well. Of course, we can also measure the $XX$ check above this box, and the box above that etc, to color an entire strip red.

In a move where a vertical strip is painted that does not include either the top or bottom box in the same column, the associated set of measurements are all the $XX$ checks within the strip, excluding the two edges at the top and bottom of this strip. If the bottom most box in that column is also included in this strip, then the $XX$ check on the lower most edge is also included in the set of measurements. If the top most box in that column is similarly included, then we also include the $XX$ check on the top most edge. We used the example of \red boxes to illustrate these moves, though the situation with \blue boxes works perfectly analogously, with the only difference between that \blue boxes can expand into horizontal, not vertical, strips in the subsequent coloring.



\begin{figure*}[!t]
    \centering
\includegraphics[width=\columnwidth]{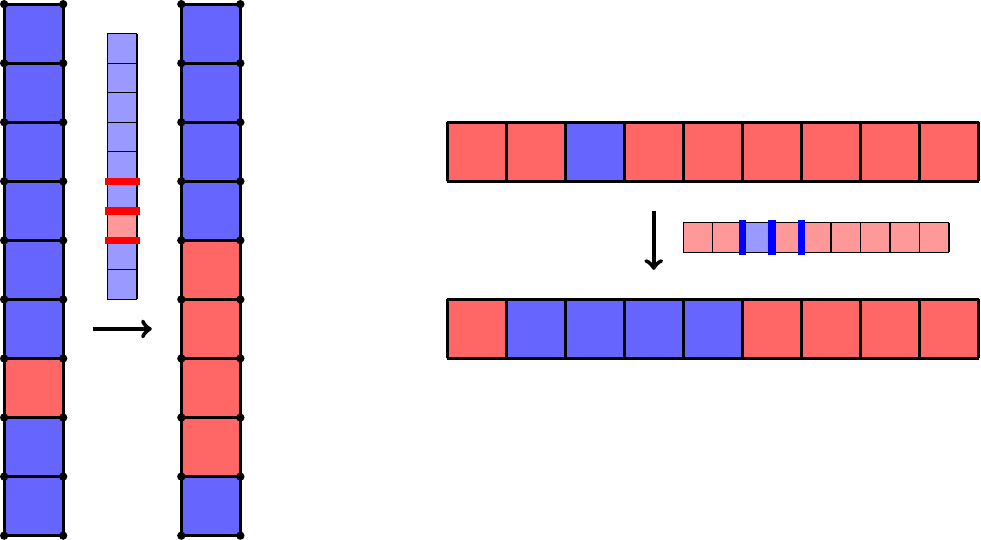}
\caption{Examples of the type of move allowed in which boxes can expand into strips in going from one coloring to the next, with all other boxes remaining unaffected. (Left) There is a single fixed $\overline{X}$ operator in the column shown. Since the product of two $\overline{X}$ operators of any two gauge qubits/boxes in the same column equals the product of $XX$ checks between those boxes, by measuring appropriate $XX$ checks along this column, we can fix additional $\overline{X}$ operators in the same column. (Right) Similarly, given a fixed $\overline{Z}$ operator in some row, we can additionally fix other $\overline{Z}$ gauge operators in the same column by measuring appropriate $ZZ$ checks.}
\label{fig:allowed-moves}
\end{figure*}

\begin{figure*}[!t]
    \centering
\includegraphics[width=\columnwidth]{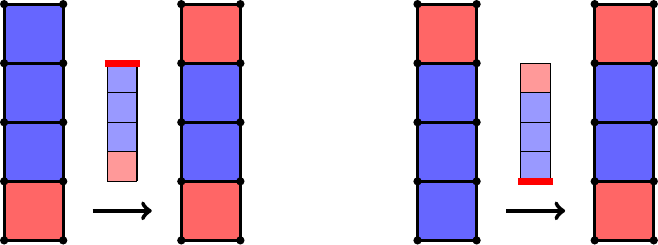}
\caption{Examples of the type of move allowed in which boxes can expand into strips that can span the lattice with periodic boundary conditions, in going from one coloring to the next, with all other boxes remaining unaffected. In both cases depicted above, the strip contains the original red box as well as another box that only vertically neighbors the original box due to the strip being allowed to extend across the lattice with periodic boundary condition. (Left) A single $XX$ check on the top edge in a column will simply measure the $\overline{X}$ operator of the gauge qubit corresponding to the top box in that column. (Right) Assuming the $X$-type stabilizer spanning this column is in the ISG, a measurement of a single $XX$ check on the bottom edge will measure the $\overline{X}$ operator of the gauge qubit corresponding to the bottom box in that column.}
\label{fig:allowed-moves-periodic}
\end{figure*}

\begin{figure*}[!t]
    \centering
\includegraphics[width=\columnwidth]{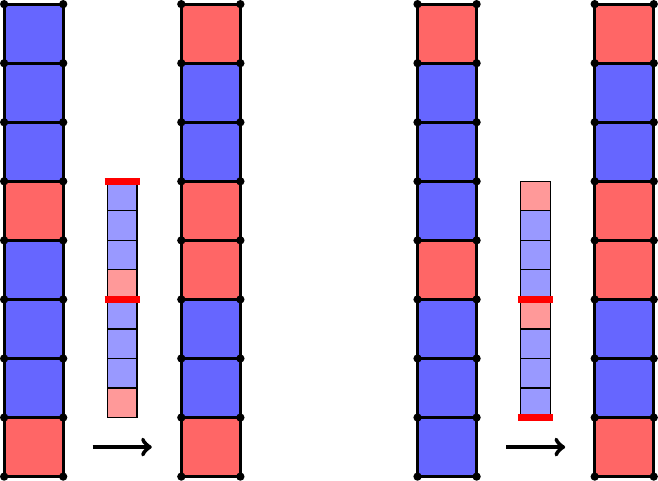}
\caption{Two patches of the coloring sequences of Fig.~\ref{fig:allowed-moves-periodic} stacked on top of another. In both cases, the top/bottom most measurement colors the top/a most box \protect\red due to the same reason as in a single patch. (Left) The $XX$ check straddling the two patches propagates the bottom box of the top patch to the top box of the bottom patch. (Right) The $XX$ check straddling the two patches propagates the top box of the bottom patch to the bottom box of the top patch.}
\label{fig:allowed-moves-periodic-patched}
\end{figure*}

\begin{figure*}[!t]
    \centering
\includegraphics[width=\linewidth]{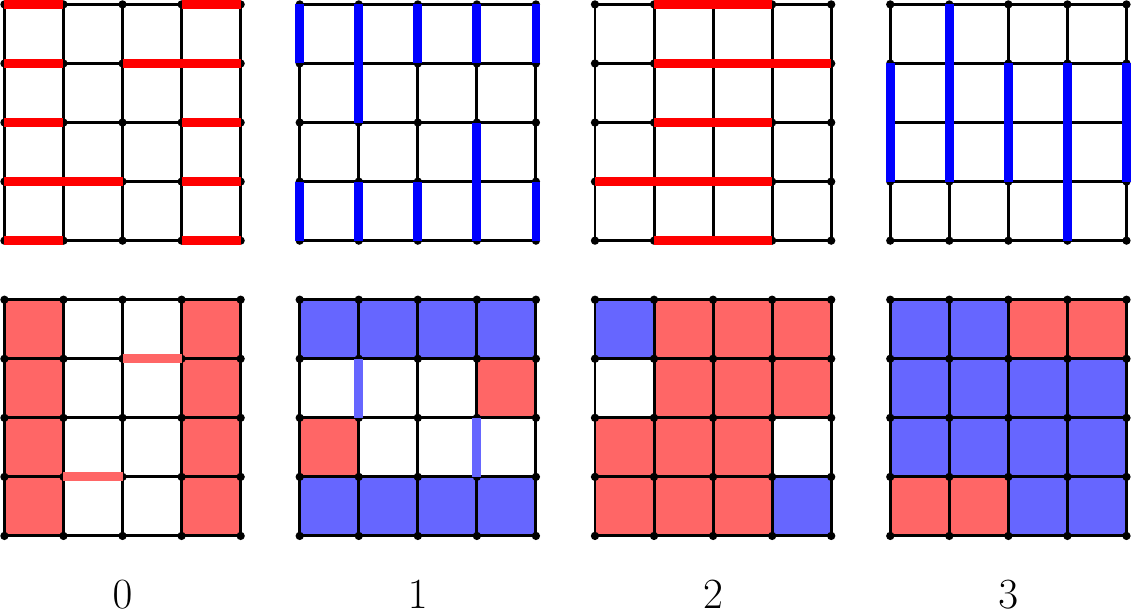}
\caption{Initial round of measurements that produce some partial colorings of the board before producing a full coloring at round 3. Thereafter, the ISGs achieve steady state and are as shown in Fig. 1 of the main text. In the above ISGs, we color some edges red or blue to indicate that the measured checks on these edges are part of the ISG.}
\label{fig:empty-to-full}
\end{figure*}

Examples of the fundamental type of move allowed in the game to produce \red or \blue strips are depicted in Fig. \ref{fig:allowed-moves}. The fact that these strips can extend with periodic boundary conditions in the coloring board game, despite being embedded in a lattice with non-periodic boundary conditions, is depicted in Fig.~\ref{fig:allowed-moves-periodic}. Coloring sequences that exploit this allowed periodicity of the strips can be patched together to build larger lattice sized solutions as shown in Fig.~\ref{fig:allowed-moves-periodic-patched}.

We now describe in detail the moves that allow us to move from one coloring to the next in the periodic sequence depicted in Fig. 1 of the main text. We denote the coloring at round $j$ as $C_j$, and describe the moves involved in going from $C_j$ to $C_{(j+1) \; \text{mod} \; 4}$ for each $j \in {0,1,2,3}$ below.

\begin{itemize}

\item $\mathbf{C_0 \rightarrow C_1}:$ We paint the entire top and bottom rows \blue in $C_1$, using the fact that there is a \blue box in each of those rows in $C_0$. This corresponds to measuring all $ZZ$ checks in the top and bottom most rows. In the remaining rows, we paint \blue horizontal strips in $C_1$ each 2 boxes long that include the single \blue box in each of those in $C_0$. This corresponds to measuring the single $ZZ$ check straddling the single edge within the strip in each of these two rows. All remaining boxes in $C_1$ inherit the same color they had in $C_0$.

\item $\mathbf{C_1 \rightarrow C_2}:$ We paint the entire middle two columns \red in $C_2$, using the fact that there is a \red box in each of those columns in $C_1$. This corresponds to measuring all $XX$ checks in these two columns. In the remaining columns, we paint \red vertical strips in $C_2$ each 2 boxes long that include the single \red box in each of those in $C_1$. This corresponds to measuring the single $XX$ check straddling the single edge within the strip in each of these two columns. All remaining boxes in $C_2$ inherit the same color they had in $C_1$.

\item $\mathbf{C_2 \rightarrow C_3}:$ We paint the entire middle two rows \blue in $C_3$, using the fact that there is a \blue box in each of those rows in $C_2$. This corresponds to measuring all $ZZ$ checks in these two rows. In the remaining rows, we paint \blue horizontal strips in $C_3$ each 2 boxes long that include the single \blue box in each of those in $C_2$. This corresponds to measuring the single $ZZ$ check straddling the single edge within the strip in each of these two rows. All remaining boxes in $C_3$ inherit the same color they had in $C_2$.

\item $\mathbf{C_3 \rightarrow C_0}:$ We paint the entire left and right most columns \red in $C_0$, using the fact that there is a \red box in each of those rows in $C_3$. This corresponds to measuring all $XX$ checks in the left and right most columns. In the remaining columns, we paint \red vertical strips in $C_0$ each 2 boxes long that include the single \red box in each of those in $C_3$. This corresponds to measuring the single $XX$ check straddling the single edge within the strip in each of these two rows. All remaining boxes in $C_0$ inherit the same color they had in $C_3$.





\end{itemize}

In Fig. \ref{fig:empty-to-full}, we also explicitly show that 
an initial single round of measurements suffices to achieve the steady state sequence of ISGs, or board colorings, shown in Fig. 1 of the main text. During the first round, not all gauge degrees of freedom are fixed, which is why we leave some boxes uncolored. However, we do color some edges red or blue, in certain cases where we just measured an $XX$ or $ZZ$ check on that edge. As noted earlier, measuring an $XX$ ($ZZ$) check on an edge may also be thought of as fixing the product of $\overline{X}$ ($\overline{Z}$) operators of neighboring boxes that share that edge.

We also note that strips can extend with periodic boundary conditions. For instance, if the top/bottom box is colored \red in some coloring, then the subsequent coloring can have the bottom/top box in the same column colored \red, so long as the top/bottom box also continues to be colored \red. This type of move is illustrated in Fig. TODO. We note that this move was not used in the $5 \times 5$ lattice solution discussed in the main text, but can be used to construct a $4 \times 4$ solution with period 6 as discussed below in Section \ref{subsecn:4x4}.




\section{Formal statements}
\label{appdx:thm-proofs}

It is helpful to first state precisely what we mean by a detector, as well as the weight of a detector. Detectors are products of measured Pauli operators that have a deterministic measurement outcome in the absence of any errors. Typically, these detectors are built out of stabilizer operators which commute with all measured operators. The product of two consecutive measurements of such stabilizers is a detector, since it evaluates to $+1$ in the absence of errors, with a measurement of $-1$ instead signaling the presence of a detectable error that anti-commutes with this stabilizer and that occurred between the two measurements.
For a subsytem code, these detectors are typically formed out of the elements of the stabilizer subgroup. In the case of the Bacon-Shor code, these stabilizers grow in weight as $O(d)$ on a $d \times d$ square lattice, and prevent the code from obtaining a threshold when used as detectors.

More generally, a detector can be described as a product of measured checks (or operators) across several rounds, such that the product of all previously measured checks (contributing to the detector) commutes with the current round, and the entire product equals a constant. Put more simply, a detector is a parity constraint on some subset of measurement outcomes that is deterministic in the absence of any errors \cite{gidney2021stim,derks2025designing}. Detectors are also sometimes referred to as ``checks"  in the literature \cite{delfosse2023spacetime}, but in this paper we use ``checks" to refer to the measured operators of the subsystem code.





The simplest choice of a detector arises when we repeatedly measure all commuting stabilizers of some stabilizer code, and form detectors out of products of successive measurements of each stabilizer. Note that for a usual detector built out of the repeated measurements of some stabilizer operator, the weight is simply twice the size of the support of the stabilizer. For instance, the usual weight-4 stabilizers in a standard surface code give rise to weight-8 detectors.

In subsystem codes, such as the Bacon-Shor code, these stabilizers are built out of the product of several checks, but the detectors are again formed from the successive measurements of such products.
However, more general detectors that do not cleanly fit the description of successive measurements of the same operator are also possible. In particular, the detectors described in this paper could be thought of as dynamical detectors, in that they do not simply measure and then re-measure the value of some static operator.

The weight of any Pauli operator is identified as the number of qubits it non-trivially acts on, while the support of a Pauli operator is the set of qubits it non-trivially acts on. For example, the weight of the 4-qubit operator $\mathbb{I} \otimes X \otimes \mathbb{I} \otimes Z$ is 2, while its support consists of the qubits that are acted on by the $X$ and $Z$ operators.
Since detectors are products of Pauli operators, the most natural definition of the weight of a detector is the sum of weights of the measured operators it is composed of. Its support can also be similarly defined as the union of supports of the measured operators it is composed of.



\begin{proposition}
A solution to the board game described in the main text on a $d \times d$ lattice with period $R$ corresponds to a measurement schedule of the Bacon-Shor check operators on a $d \times d$ lattice with period $R$, that produces a steady state ISG sequence corresponding to the coloring sequence, and produces detectors of weight $O(Rd)$.
\label{thm:coloring-ISG}
\end{proposition}
\begin{proof}[\textbf{Proof.}]
First, we describe how to extract the measurement schedule from the coloring sequence that solves the game. The colorings in the sequence correspond to the ISGs once they have achieved steady state. Each \red (\blue) box is a gauge fixed $\overline{X}$ ($\overline{Z}$) operator. In going from one coloring to the next in the sequence, we translate the allowed moves as follows.
For a move that produces a vertical (horizontal) \red (\blue) strip that excludes both the top and bottom most box, we measure the $XX$ ($ZZ$) checks on all the edges contained within the vertical (horizontal) strip produced, excluding the two edges on either end of this strip. If the strip includes the top (right) most box in that column (row), then we also include the $XX$ ($ZZ$) check on the top (right) most edge of the strip. Similarly, if the strip includes the bottom (left) most box in that column (row), then we also include the $XX$ ($ZZ$) check on the bottom (left) most edge of the strip. This provides the measurement schedule that produces the periodic sequence of ISGs in steady state. The allowed moves and the measurements that enable them are depicted in Figs. \ref{fig:basic-moves} and \ref{fig:allowed-moves}.\\

Next, we bound the weight of all detectors produced by this measurement schedule. If there is one \red (\blue) box in a column (row), there are two associated elements in the ISG: the product of all $XX$ ($ZZ$) checks above (to the right of) it, and product of all $XX$ ($ZZ$) checks below (to the left of) it. These can each be used as detectors if the entire column's (row's) $XX$ ($ZZ$) checks are measured in the next round, so that all boxes in the column (row) are colored \red (\blue). If there are several \red (\blue) boxes in a column (row), then the product of $XX$ ($ZZ$) checks between closest \red (\blue) boxes within the column (row) are also part of the ISG, and can be used as detectors if measured. In all cases, detectors are only ever formed out of the product of checks within the same column (row). In each column (row), there are at most $d$ checks being measured in each of $R$ rounds. Since the detector is a subset of all such measured checks, the number of measured checks appearing in the detector, and therefore the weight of the detector, is at most $O(Rd)$.
\end{proof}

\begin{proposition}
Given a solution to the board game for a $d \times d$ lattice in the form of a periodic coloring sequence with period $R$, there exists a solution for any $L \times K$ lattice, where $L,K \geq d$, in the form of a similar periodic coloring sequence with period $R$, producing detectors of weight $O(Rd)$. In particular, if a weight of any detector in the $d \times d$ solution is at most $w$, then the weight of any detector in the generalized solution is at most $2(d + w - 1)$.
\label{thm:generalization}
\end{proposition}
\begin{proof}[\textbf{Proof.}]
First, we note that for the solution to the $d \times d$ lattice, that we refer to as the sub-solution here, all detectors have weight at most $w \in O(Rd)$ as noted in Theorem \ref{thm:coloring-ISG}. 
Next, a generalization to any square lattice of size $L \times L$, where $L = n(d-1) + 1$ and $n$ is a positive integer, can be constructed by simply stacking together the colorings in the sequence of the sub-solution. More formally, if $A^{(r)}$ is a $(d-1) \times (d-1)$ matrix with entries $A^{(r)}_{ij} \in \{\red, \blue \}$ representing the coloring, or ISG, at round $r$,
then the corresponding coloring, or ISG, for an $L \times L$ lattice with $L = n(d-1) + 1$, is given by
\begin{equation}
A_{n}^{(r)} = 
J_{n-1} \otimes A^{(r)}
\end{equation}
where $J_n$ is the $n \times n$ matrix with all entries equaling 1, and $J_0 = 1$. This produces a stacking such that there are $n$ stacked copies of the sub-solution in both the verical and horizontal directions. In such a coloring, besides the detectors being produced in each stack given by the sub-solution, there are additional detectors formed out of the measured checks between a \red (\blue) box from one copy of the sub-solution and its neighboring \red (\blue) box in the same column (row) in a vertically (horizontally) neighboring copy.
The distance, captured by the number of edges and therefore the number of measured checks, between these two boxes is at most $d-1$. Since this is true of all colorings in the sequence, and there are $R$ such colorings, the detectors that are formed between these neighboring boxes consist of at most $O(Rd)$ many checks, which therefore sets an upper bound to the weight of the detectors thus formed. To obtain a solution for a non-square lattice $L \times K$, where $K < L$, we simply truncate the pattern thus obtained to the desired size. Note that we require $L,K \geq d$. In particular, truncating a given solution to a smaller board size may not yield a board game solution, so that $L,K < d$ may not always lead to a solution.

We can be even more precise in relating the maximum detector weight in the sub-solution to that in the generalized case. Suppose we have a detector formed out of the top most $\red$ box in a column in the sub-solution, given by $D = \prod_{r=1}^{R} \left( \prod_{j=1}^{M_r} m_{j}^{(r)} \right)$. Its weight is given by
\begin{eqnarray}
\texttt{wt}(D) &=& \sum_{j=1}^{M_1} \texttt{wt}(m_{j}^{(1)}) + \sum_{r=2}^{R-1} \left( \sum_{j=1}^{M_r} \texttt{wt}(m_{j}^{(r)})\right) \nonumber \\
&& \; + \sum_{j=1}^{M_K} \texttt{wt} (m_{j}^{(K)}) \; \leq\;  w
\end{eqnarray}
The initial round 1 is when the detector is initialized in the form of some fixed \red operator in a column. The final round $R$ is when some (possibly different) \red operator in the same column is measured, producing a detector. In the sub-solution, there are exactly as many checks being measured as there are edges between these boxes and the upper lattice boundary. For the detector formed out of these boxes in the generalized case, that we denote $D'$, there are now $d-1$ many edges between the boxes in one stack and the corresponding ones in the stack above. For the intermediate rounds, where only a subset of checks in the same column (and importantly, within the same sub-solution) are being measured, there are twice the number of checks being measured for the generalized detector $D'$. Thus, we have
\begin{eqnarray}
\texttt{wt}(D') &\leq& (d-1) + 2 \sum_{r=2}^{R-1} \left( \sum_{j=1}^{M_r} \texttt{wt}(m_{j}^{(r)})\right) + (d-1) \nonumber \\
&\leq& 2 (d + w - 1)
\end{eqnarray}
The above argument can be straightforwardly generalized to the case of neighboring \blue boxes in the same row across neighboring copies of the sub-solution.
\end{proof}

\section{Detector examples}
\label{appdx:detector-examples}

\begin{figure*}[!h]
    \centering
\includegraphics[width=\linewidth]{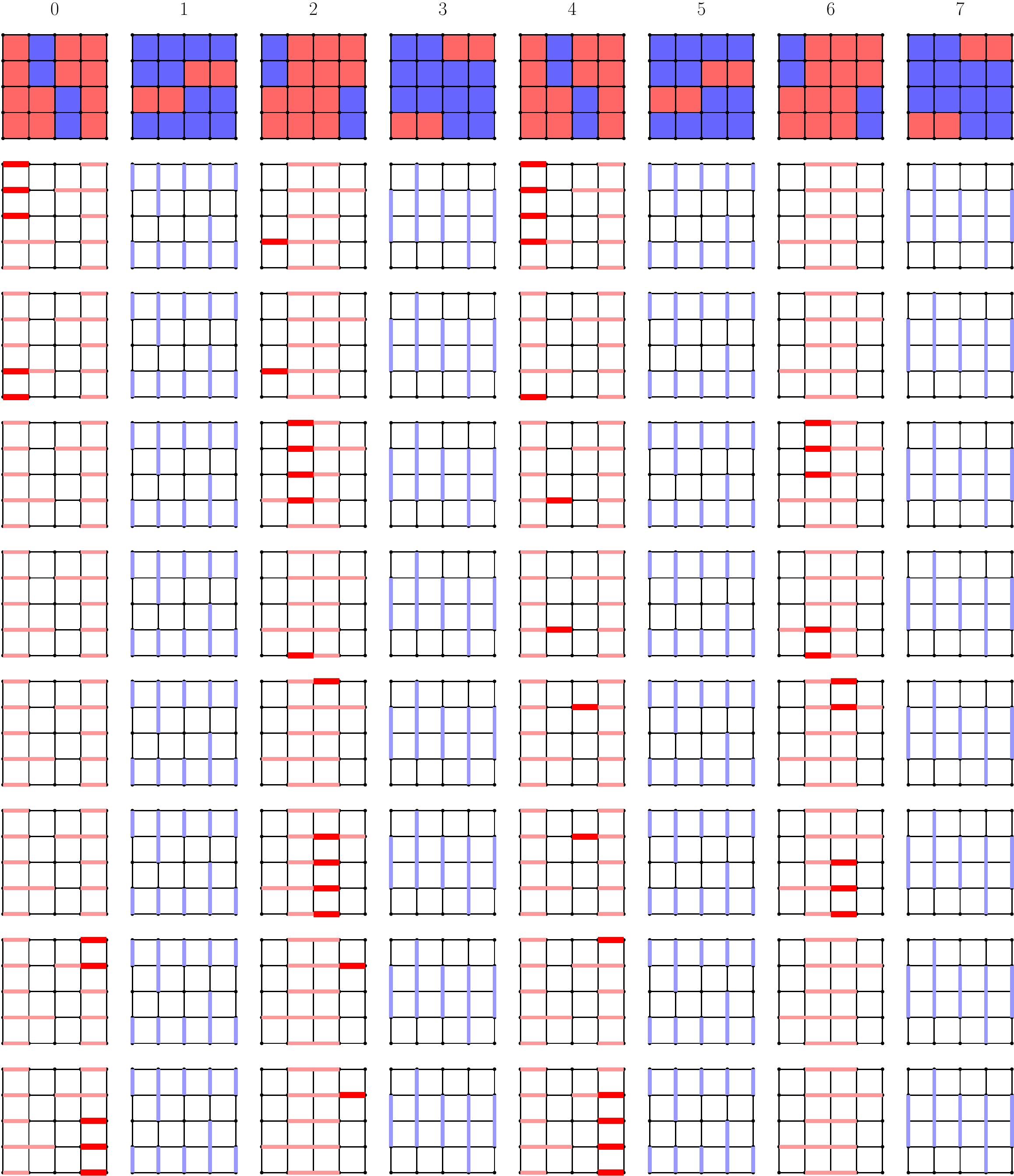}
\caption{All $X$-type detectors on a $5 \times 5$ square lattice. In each row of grids, the highlighted red checks represent the measurement whose product deterministically equals $+1$ in the absence of any errors.}
\label{fig:detectors-X-4x4}
\end{figure*}

\begin{figure*}[!h]
    \centering
\includegraphics[width=\linewidth]{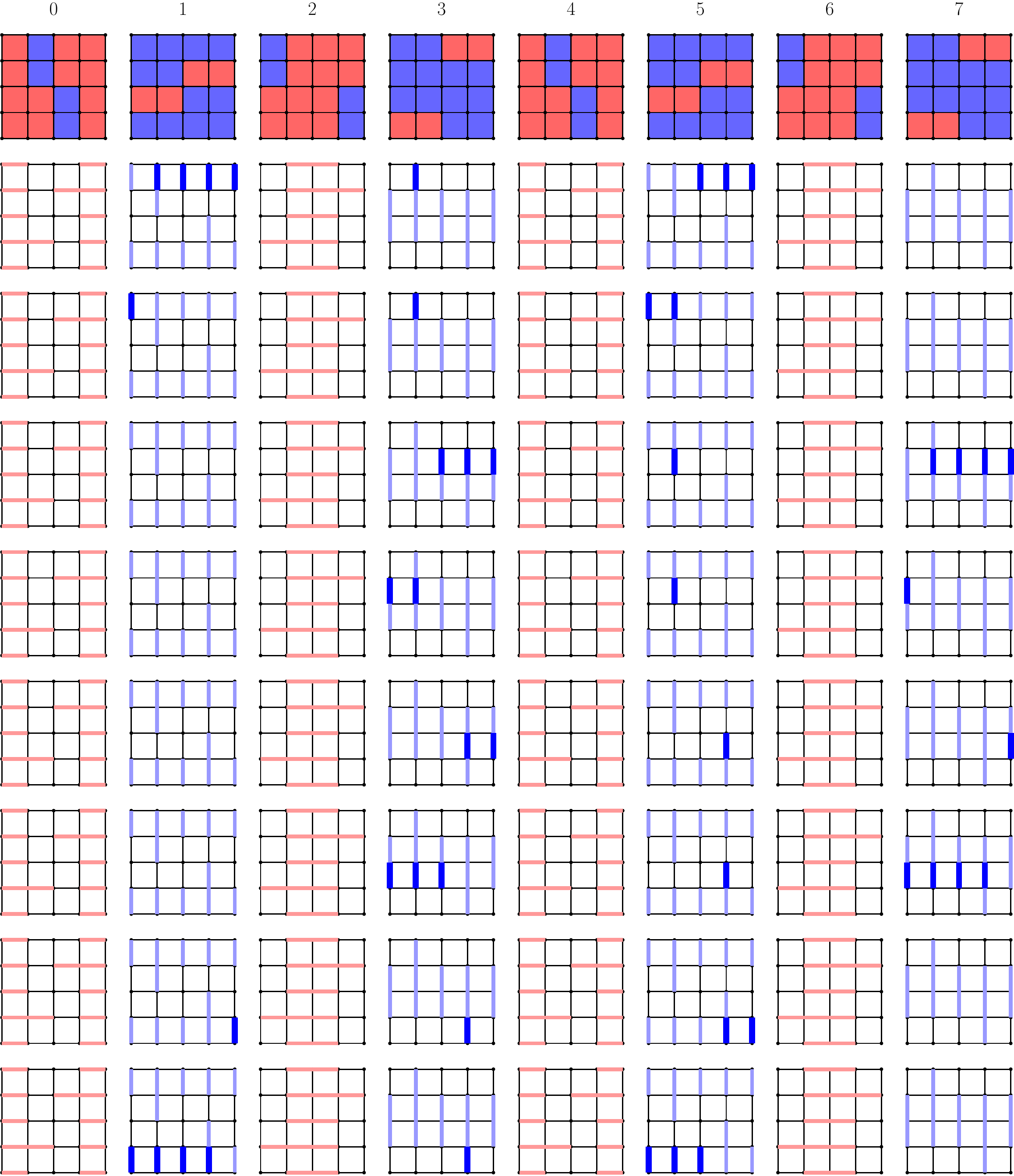}
\caption{All $Z$-type detectors on a $5 \times 5$ square lattice. In each row of grids, the highlighted blue checks represent the measurement whose product deterministically equals $+1$ in the absence of any errors.}
\label{fig:detectors-Z-4x4}
\end{figure*}

\begin{figure*}[!h]
    \centering
\includegraphics[width=\linewidth]{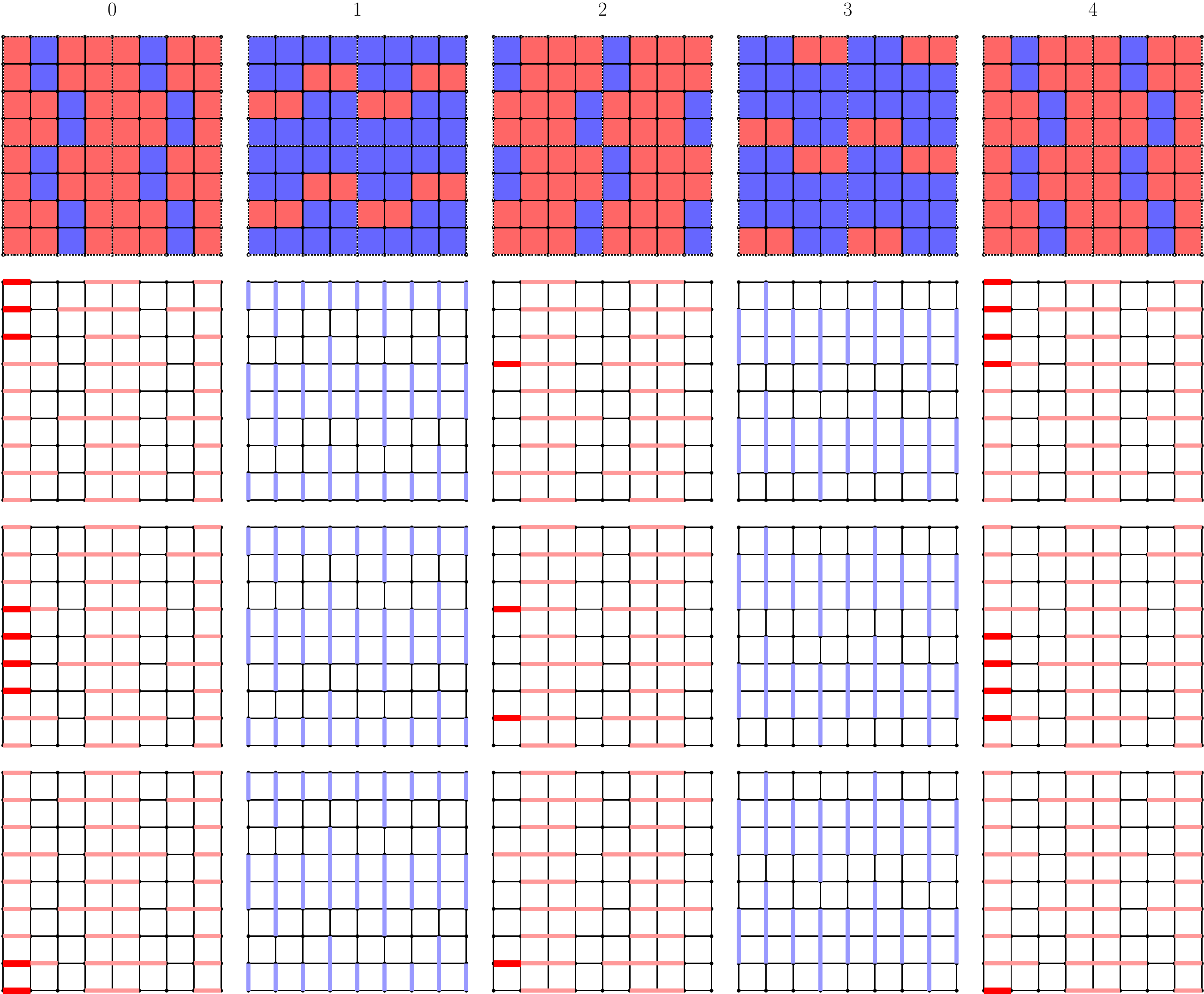}
\caption{All three $X$-type detectors in the left most column of the $9 \times 9$ lattice solution, built out of the smaller $5 \times 5$ solution, which are each drawn inside of the white dotted lines. The largest weight of any detector among these is the one that cuts across two sub-solutions. This detector is composed of 10 weight-2 checks, and therefore has weight 20. Although we only considered $X$-type detectors in the left most column, the above argument works exactly analogously for other columns, and with all rows and their associated $Z$-type detectors.}
\label{fig:detectors-9x9}
\end{figure*}

Here, we explicitly describe all detectors for the Bacon-Shor code on a $5 \times 5$ square lattice with the period-4 measurement schedule described in the main text. All $X$-type detectors are shown in Fig. \ref{fig:detectors-X-4x4} while all $Z$-type detectors are shown in Fig. \ref{fig:detectors-Z-4x4}. For the $9 \times 9$ generalization depicted in Fig. 2 of the main text, we show all detectors produced in the left most column in Fig. \ref{fig:detectors-9x9}.

\begin{figure*}
\centering
\begin{subfigure}{0.3\textwidth}
    \includegraphics[width=\textwidth]{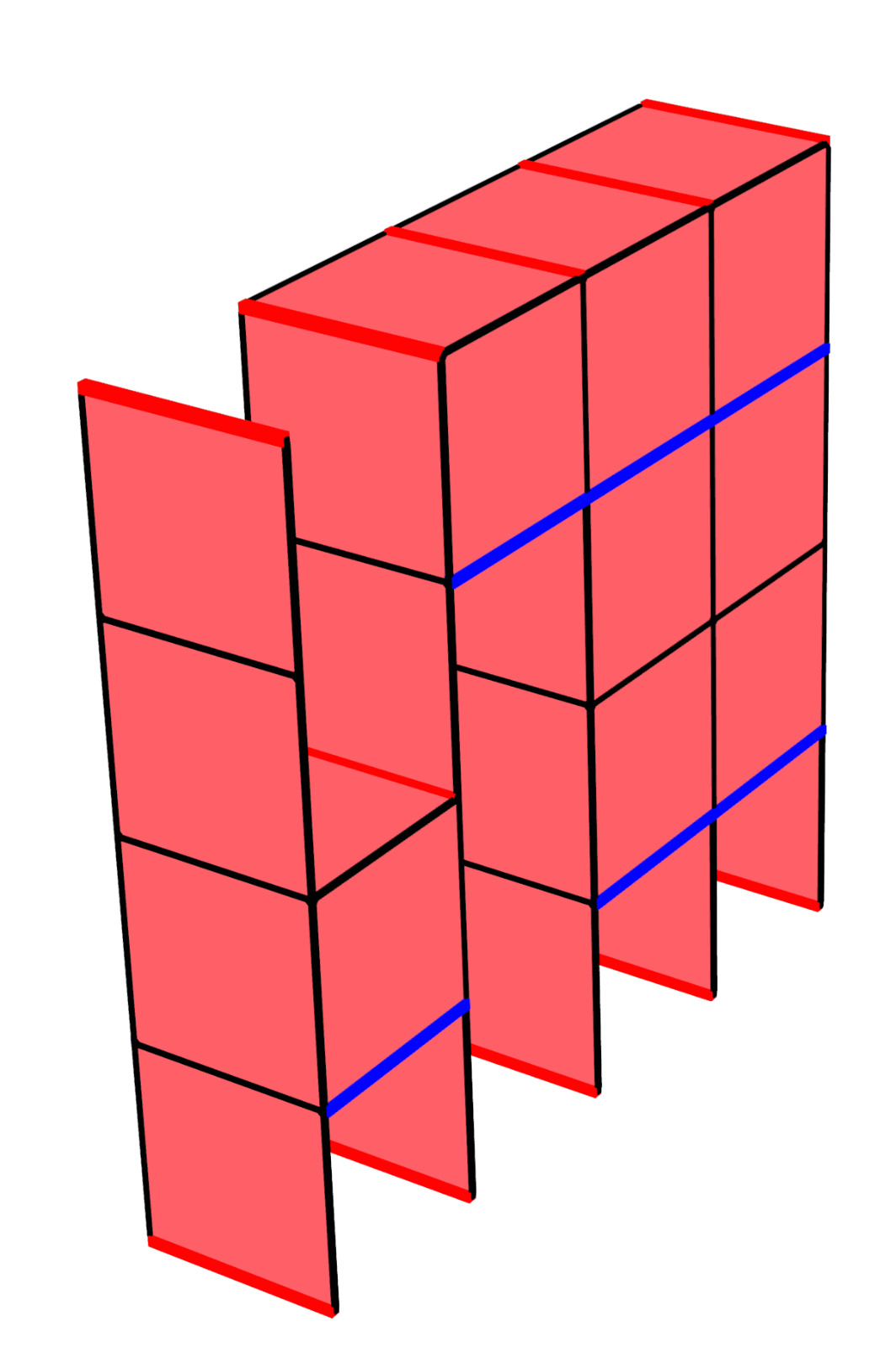}
    \caption{}
    \label{subfig:detector-3d}
\end{subfigure}
\hfill
\begin{subfigure}{0.3\textwidth}
    \includegraphics[width=\textwidth]{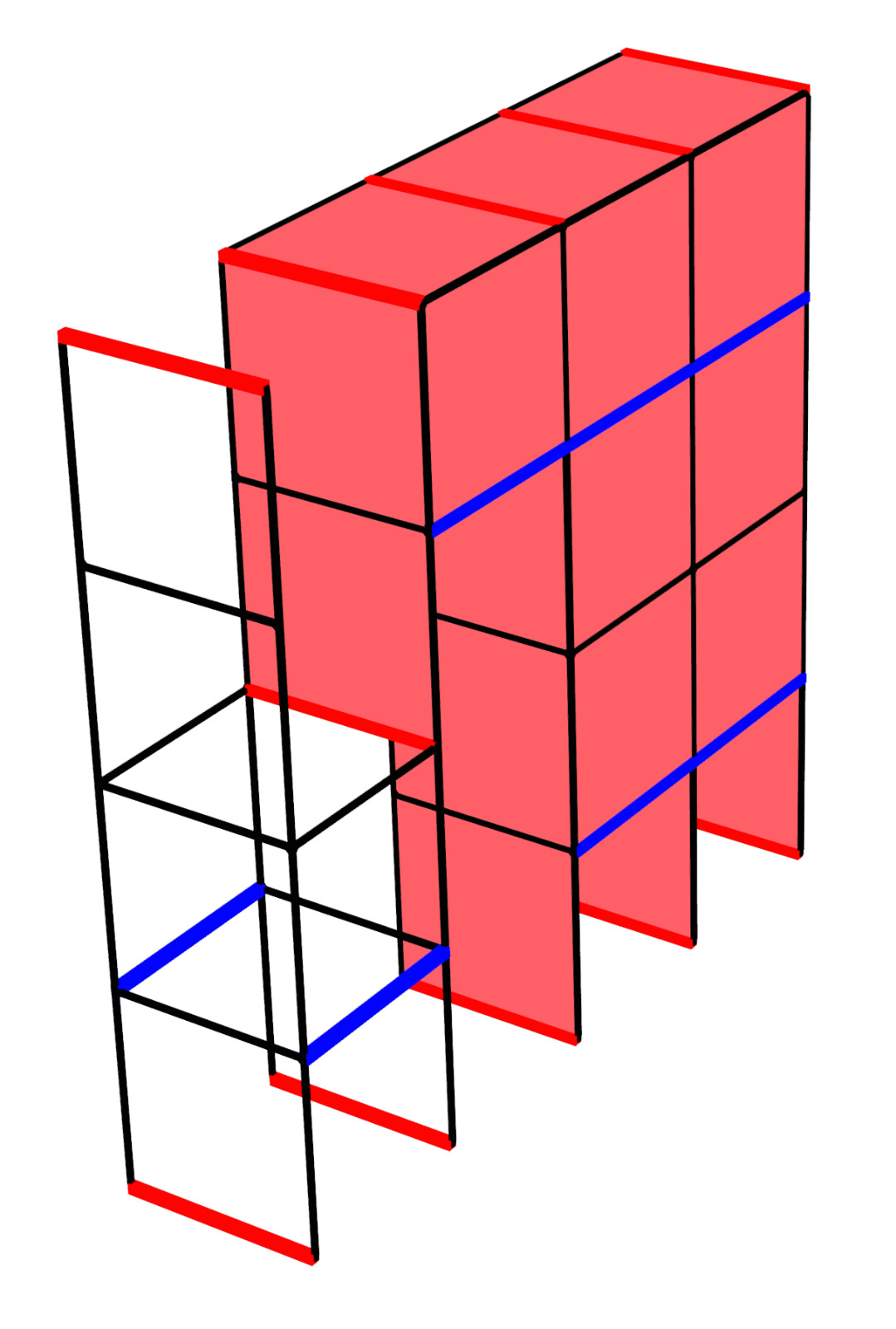}
    \caption{}
\end{subfigure}
\hfill
\begin{subfigure}{0.3\textwidth}
    \includegraphics[width=\textwidth]{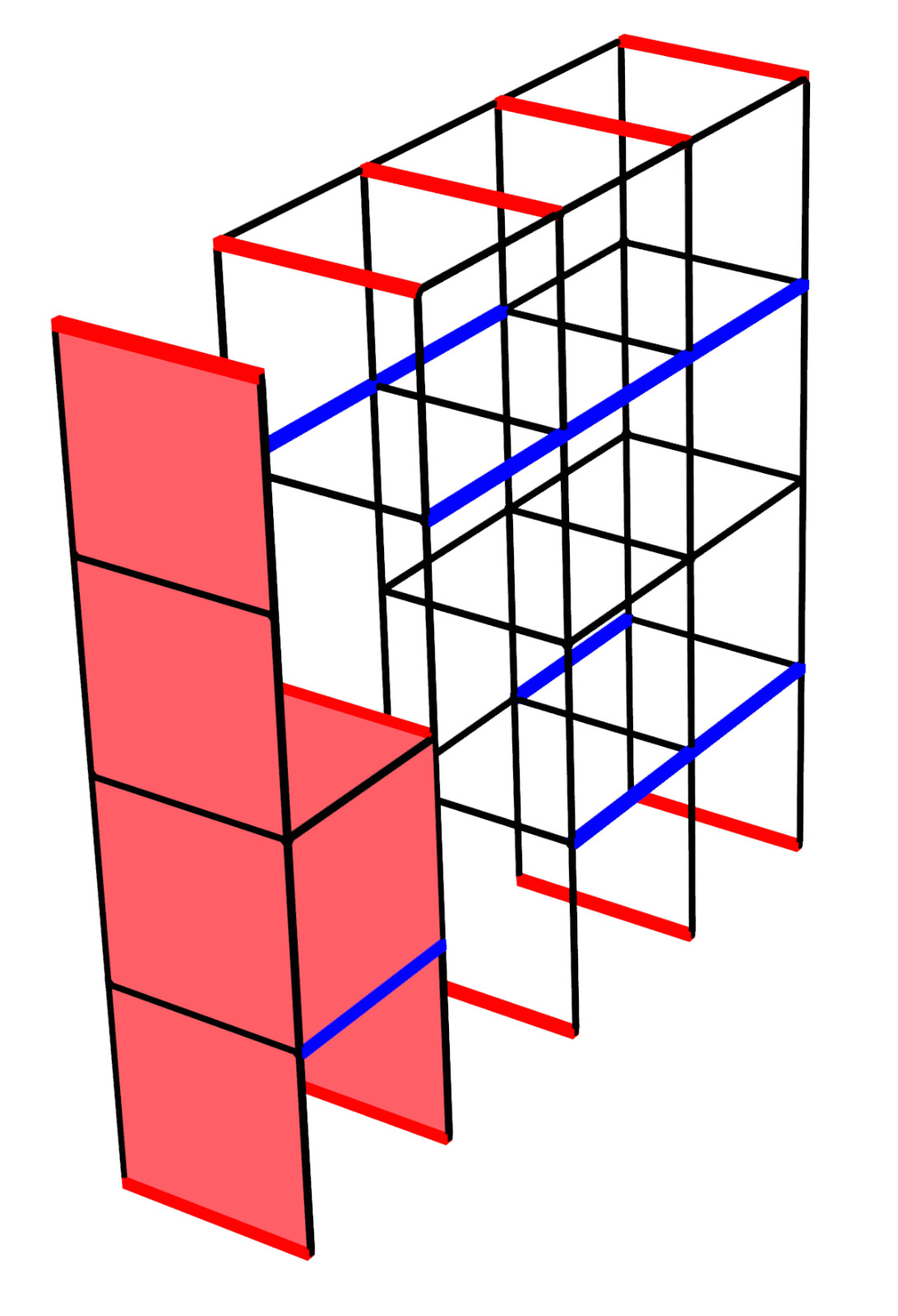}
    \caption{}
\end{subfigure}
\caption{(a) Evolution of the ISG in the leftmost column of \cref{fig:empty-to-full} over period of the schedule. Time progresses along the vertical axis and red and blue solid lines show $X$ and $Z$ measurements at specific timesteps. Red surfaces/volumes show the evolution over time of $X$ operators introduced into the ISG by these measurements. (b) and (c) Subsets of measurements that form detectors, interpretable as regions of spacetime bounded by measurements.}
\label{fig:detectors-3d}
\end{figure*}

\section{Noise model}
\label{appdx:noise}

All circuits in this paper were simulated using the uniform noise model summarized in Table~\ref{tab:noise-channels} and Table~\ref{tab:ideal-noisy-gates}. Table~\ref{tab:noise-channels} defines the individual noise channels (\(\text{MERR}_B\), \(\text{XERR}(p)\), \(\text{ZERR}(p)\), \(\text{DEP1}(p)\), \(\text{DEP2}(p)\)) used to build up noisy versions of gates. Table~\ref{tab:ideal-noisy-gates} then shows how each ideal gate is replaced by its noisy counterpart under this model. 

\begin{table}[ht]
    \centering
    \caption{Definitions of noise channels used to define noisy versions of gates in Table~\ref{tab:ideal-noisy-gates}.}
    \label{tab:noise-channels}
    \renewcommand{\arraystretch}{1.2}  
    \begin{tabular}{l p{5cm}}
    \hline
    \textbf{Noise Channel} & \textbf{Probability Distribution of Effects} \\
    \hline
    \(\text{MERR}_B(p)\) & 
    \((1-p) \;\rightarrow\; M_B \) \newline
    \(\quad p \;\rightarrow\; M_{-1 \cdot B}\) (i.e., measurement result is inverted) \\
    \hline
    \(\text{XERR}(p)\) & 
    \((1-p) \;\rightarrow\; I \) \newline
    \(\quad p\;\rightarrow\; X \) \\
    \hline
    \(\text{ZERR}(p)\) & 
    \((1-p) \;\rightarrow\; I \) \newline
    \(\quad p\;\rightarrow\; Z \) \\
    \hline
    \(\text{DEP1}(p)\) & 
    \( (1-p) \;\rightarrow\; I \) \newline
    \(\quad \tfrac{p}{3} \;\rightarrow\; X, \quad \tfrac{p}{3} \;\rightarrow\; Y, \quad \tfrac{p}{3} \;\rightarrow\; Z \) \\
    \hline
    \(\text{DEP2}(p)\) & 
    \( (1-p) \;\rightarrow\; I \otimes I\) \newline
    \(\quad \tfrac{p}{15} \;\rightarrow\; I \otimes X, \; I \otimes Y, \; I \otimes Z, \; X \otimes I; X \otimes X; X \otimes Y; X \otimes Z; Y \otimes I; Y \otimes X; Y \otimes Y; Y \otimes Z; Z \otimes I; Z \otimes X; Z \otimes Y; Z \otimes Z\)\newline
    \textit{(16 equally likely two-qubit error outcomes, each with probability }p/15\textit{)} \\
    \hline
    \end{tabular}
\end{table}

\begin{table}[ht]
    \centering
    \caption{Uniform noise model used in all simulations. Each ideal gate (left column) is replaced by the corresponding noisy gate (right column). Noise channels are defined in Table~\ref{tab:noise-channels}.}
    \label{tab:ideal-noisy-gates}
    \renewcommand{\arraystretch}{1.2}
    \begin{tabular}{l l}
    \hline
    \textbf{Ideal Gate} & \textbf{Noisy Gate} \\
    \hline
    Idle       & \(\text{DEP1}(p)\) \\
    \(R_X\)    & \(\text{ZERR}(p) \cdot R_X\) \\
    \(R_Z\)    & \(\text{XERR}(p) \cdot R_Z\) \\
    \(M_X\)    & \(\text{DEP1}(p) \cdot \text{MERR}_X(p)\) \\
    \(M_Z\)    & \(\text{DEP1}(p) \cdot \text{MERR}_Z(p)\) \\
    \(M_{XX}\) & \(\text{DEP2}(p) \cdot \text{MERR}_{XX}(p)\) \\
    \(M_{ZZ}\) & \(\text{DEP2}(p) \cdot \text{MERR}_{ZZ}(p)\) \\
    \hline
    \end{tabular}
\end{table}

Note that \(\text{MERR}_{X}(p)\) and \(\text{MERR}_{Z}(p)\) flip measurement outcomes of single-qubit \(X\)- or \(Z\)-type measurements with probability \(p\), while \(\text{MERR}_{XX}(p)\) and \(\text{MERR}_{ZZ}(p)\) do the same for two-qubit \(XX\)- or \(ZZ\)-type measurements.

\section{Other possible moves}
The basic moves we allowed for are sufficient to solve the board game, and thereby lead to constant weight detectors, as shown earlier. We note here however that these are not the only possible moves. In particular, if the bottom box in some column is painted \red, then in the subsequent coloring, the top box in the same column can also be colored \red as long as the bottom box in the same column continues to be colored \red.

\begin{figure*}[!h]
    \centering
\includegraphics[scale=1.0]{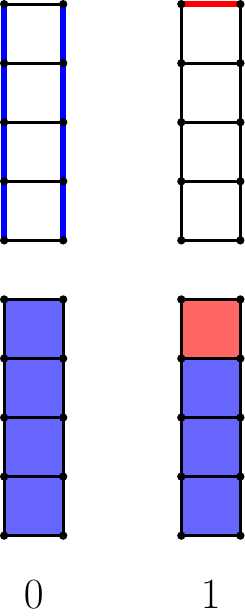}
\caption{An example of a move made possible through a measurement schedule of the Bacon-Shor check operators, but that, for simplicity, is disallowed in the rules of the board game as described in the main text. In the $0$th round, we measure all vertical $ZZ$ checks on a $5 \times 2$ lattice, then simply measure a single $XX$ check on the top-most edge.}
\label{fig:disallowed-moves}
\end{figure*}

For any fixed size square lattice, the top box in any column can be painted \red simply by measuring the $XX$ check on the top of this box. Coloring the top box in a column \red while keeping the bottom box in the same column \blue is depicted in Fig~\ref{fig:disallowed-moves}.

\begin{figure*}[!h]
    \centering
\includegraphics[scale=1.0]{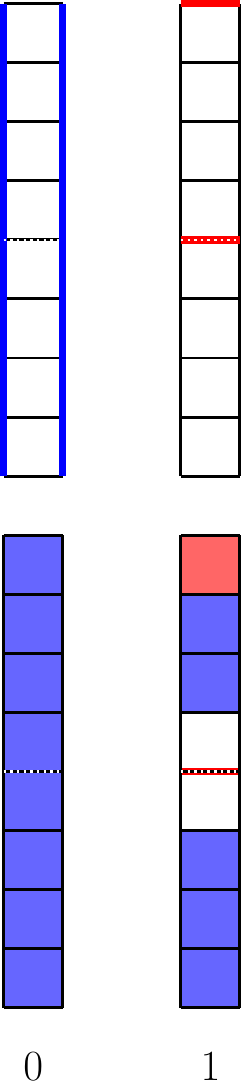}
\caption{Two patches of the coloring sequence from Fig. \ref{fig:disallowed-moves}, stacked one on top of the other. Note that neither the top box in the bottom patch nor the bottom box in the top patch can be colored red with this schedule. Only the product of the logical $\overline{X}$ operators of those two gauge qubits exists in the ISG in the $1$-st round. For this reason, we disallow such moves in the board game.}
\label{fig:disallowed-moves-patched}
\end{figure*}

The reason we would require the bottom most box to also be painted \red is to allow the patching of this fixed size lattice to build solutions for larger lattices. The same kind of move can be replicated on larger sized lattices by patching together the ISGs and then inferring the measurement schedule in a manner similar to that used to obtain the board game solution in the main text. This is possible only because the bottom box in some column of some patch is painted \red. The bottom edge of this box is the top edge (in that column) of the patch directly below it. A single $XX$ check on this edge therefore propagates the \red box one unit below, which appears as the top box in the patch below.
If only the top, but not the bottom, box in a column gets colored \red, then such a move does not patch together to build larger solutions, as depicted in Fig. \ref{fig:disallowed-moves-patched}.

In all of the above, the same holds true if we interchange the role of the top and the bottom box. The same also holds true if we replace
\begin{itemize}
\item \red with \blue,
\item `column' with `row',
\item $XX$ with $ZZ$,
\item `top' with `right'/`left',
\item and `bottom' with (respectively) `left'/`right'.
\end{itemize}

\section{Smaller board sizes}
It is natural to wonder whether board game solutions exist for board sizes smaller than the one described on a $5 \times 5$ lattice in the main text. First, we argue that it is impossible to find a solution to the board game for a $3 \times 3$ lattice. We then also show that a period 6 schedule exists for a $4 \times 4$ lattice.

\subsection{$3 \times 3$ lattice}

\begin{figure*}[!h]
    \centering
\includegraphics[scale=1.0]{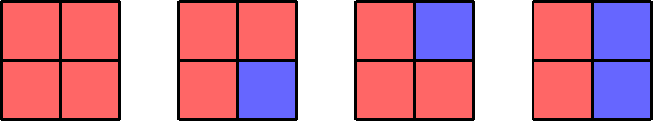}
\caption{Each of the possible colorings of a $3 \times 3$ lattice with the left-most column colored red violates rule 2 of the board game. Therefore, there is no $3 \times 3$ solution to the board game.}
\label{fig:no-soln-3x3}
\end{figure*}

By rule 1 of the board game from the main text, each plaquette column (row) is colored entirely red (blue) in at least one coloring. Let us consider a coloring of a $3 \times 3$ lattice where the left plaquette column is painted entirely red. This leaves four possibilities for the coloring of the plaquette column to the right, as depicted in Fig.~\ref{fig:no-soln-3x3}.

Each of these possibilities violates Rule 2 of the board game, which states that each column (row) contains at least one red (blue) plaquette in each coloring. However, they do so for different reasons. Considering each of the colorings in Fig.~\ref{fig:no-soln-3x3} from L-R, the colorings violate rule 2 because:
\begin{itemize}
\item (a) There is no row that contains even a single blue plaquette;
\item (b) The top row does not contain even a single blue plaquette;
\item (c) The bottom row does not contain even a single blue plaquette;
\item (d) The right column does not contain even a single red plaquette.
\end{itemize}

\subsection{$4 \times 4$ lattice}
\label{subsecn:4x4}

\begin{figure*}[!h]
    \centering
\includegraphics[width=\textwidth]{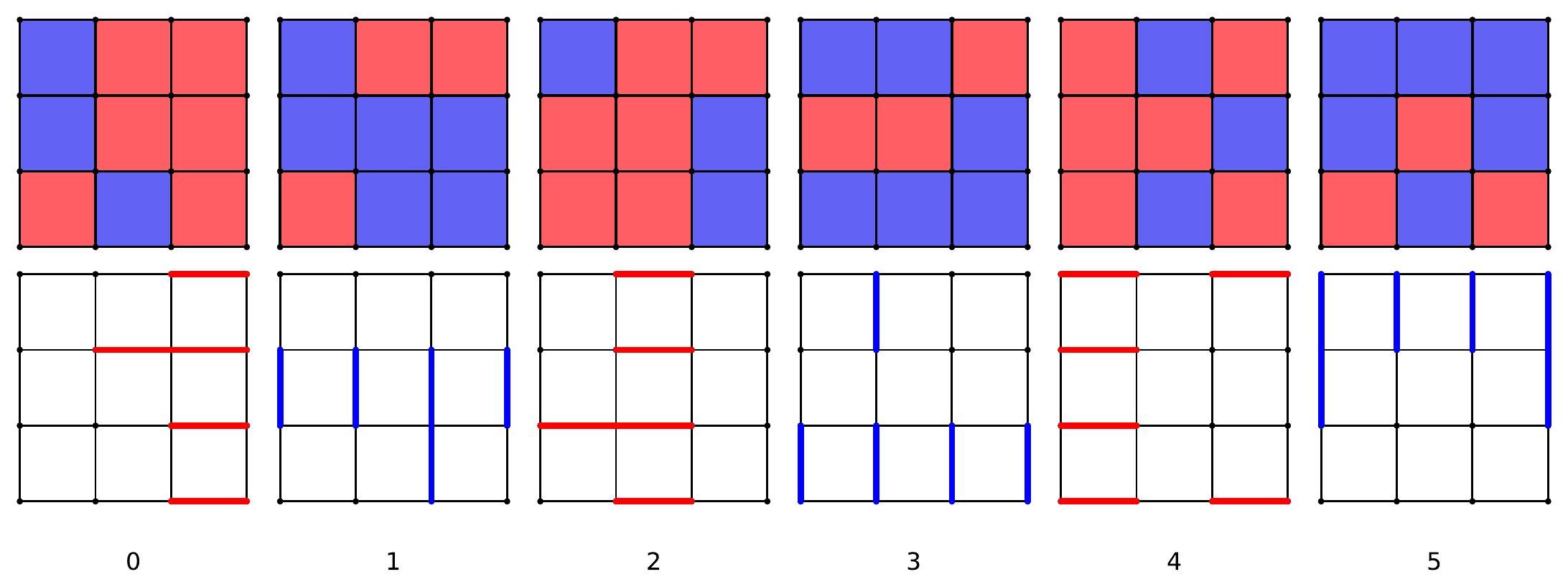}
\caption{A solution to the board game on a $4 \times 4$ lattice. }
\label{fig:4x4_lattice_soln_2}
\end{figure*}

Unlike the $3 \times 3$ case, a valid solution to the board game does exist on a $4 \times 4$ lattice. We found a periodic coloring sequence with period $6$ that satisfies all rules of the board game. In particular, partly exploits the fact that colored strips can extend across the lattice with periodic boundary conditions as shown in Figs.~\ref{fig:allowed-moves-periodic} and \ref{fig:allowed-moves-periodic-patched}. The corresponding sequence of colorings is shown in Fig.~\ref{fig:4x4_lattice_soln_2}.

This establishes that the smallest board size admitting a valid solution is $4 \times 4$. As described in Proposition~\ref{thm:coloring-ISG}, the coloring sequence can be translated directly into a measurement schedule of the Bacon–Shor check operators, producing a periodic steady-state ISG sequence with constant-weight detectors. Furthermore, by Theorem~\ref{thm:generalization}, this solution can be used as a building block for constructing schedules on larger lattices while preserving constant detector weight.

The present construction has a maximum detector weight of $16$, which is smaller than that of the $5 \times 5$ schedule. However, it requires a longer schedule period. Over a fixed number of measurement steps, a period-$6$ schedule produces each detector less frequently than a period-$4$ schedule. From the $5 \times 5$ lattice onward, the board-game construction admits solutions with period $4$, making the $5 \times 5$ schedule more representative of the construction on larger lattices. For this reason, the $5 \times 5$ schedule is used throughout the main text as the representative example for illustrating the board game and its associated measurement schedule, rather than as a schedule with less detector weight.

\begin{figure*}[!h]
    \centering    \includegraphics[width=\linewidth]{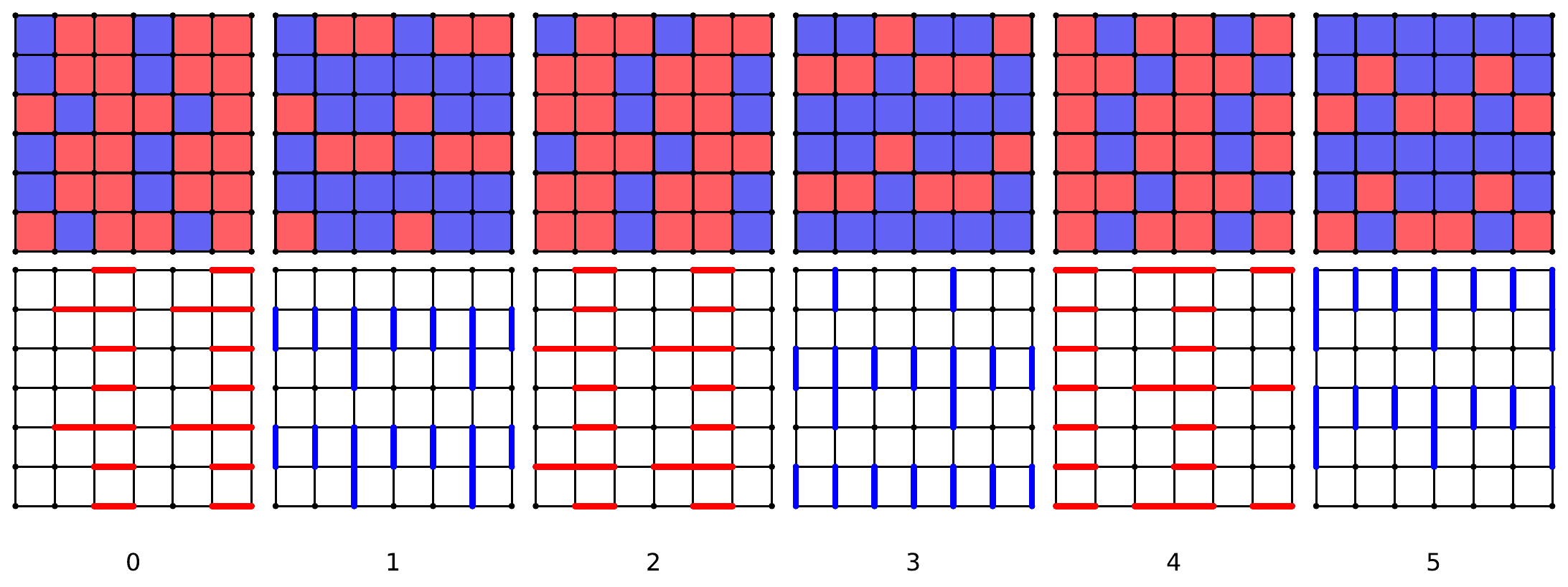}
    \caption{A solution to the board game on a $7 \times 7$ lattice, build out of the smaller $4 \times 4$ solution in \cref{fig:4x4_lattice_soln_2}.}
    \label{fig:7x7_lattice_soln_2}
\end{figure*}

\bibliography{references}